\documentclass[groupedaddress, superscriptaddress, twocolumn, pra,nofootinbib]{revtex4-2}
\usepackage{graphicx}
\usepackage{calc}
\usepackage{amsmath}
\usepackage{amsthm}
\usepackage{amssymb}
\usepackage{bm}
\usepackage{color}
\usepackage[pdftex,dvipsnames,usenames]{xcolor}
\usepackage[colorlinks=true,urlcolor=blue,citecolor=blue,linkcolor=blue]{hyperref}

\newtheorem{statement}{Statement}

\DeclareMathOperator{\Tr}{Tr}

\DeclareMathOperator{\arcosh}{arcosh}
\renewcommand{\Im}{\mbox{\rm Im}}
\renewcommand{\Re}{\mbox{\rm Re}}
\newcommand{\ket}[1]{\left|#1\right\rangle}
\newcommand{\bra}[1]{\left\langle#1\right|}
\DeclareMathOperator{\D}{d}

\usepackage{amsfonts}

\begin{document}
    \title{Nonclassical correlations of radiation in relation to Bell nonlocality}
	
	\author{V. S. Kovtoniuk}
	\affiliation{Bogolyubov Institute for Theoretical Physics, NAS of Ukraine, Vul. Metrologichna 14b, 03143 Kyiv, Ukraine}
	\affiliation{Physics Department, Taras Shevchenko National University of Kyiv, Prospect Glushkova 2, 03022 Kyiv, Ukraine}
	
	\author{I. S. Yeremenko}
	\affiliation{Bogolyubov Institute for Theoretical Physics, NAS of Ukraine, Vul. Metrologichna 14b, 03143 Kyiv, Ukraine}
	\affiliation{Physics Department, Taras Shevchenko National University of Kyiv, Prospect Glushkova 2, 03022 Kyiv, Ukraine}

	\author{S. Ryl}
	\affiliation{Institut f\"ur Physik, Universit\"at Rostock, Albert-Einstein-Stra\ss{}e 23, D-18059 Rostock, Germany}
	
	\author{W. Vogel}	
	\affiliation{Institut f\"ur Physik, Universit\"at Rostock, Albert-Einstein-Stra\ss{}e 23, D-18059 Rostock, Germany}
	
	\author{A. A. Semenov}
	\affiliation{Bogolyubov Institute for Theoretical Physics, NAS of Ukraine, Vul. Metrologichna 14b, 03143 Kyiv, Ukraine}
	\affiliation{Kyiv Academic University, Blvd. Vernadskogo  36, 03142  Kyiv, Ukraine}
	\affiliation{Institute of Physics, NAS of Ukraine, Prospect Nauky 46, 03028 Kyiv, Ukraine}
	
\begin{abstract}
	We analyze nonclassical correlations between outcomes of measurements conducted on two spatial radiation modes.
	These correlations cannot be simulated with statistical mixtures of coherent states or, more generally, with non-negative phase-space functions of quantum states and measurements.
	We argue that nonclassical correlations are naturally related to Bell nonlocality, the former being a more general class of quantum correlations. 
	Indeed, it is known that local realistic as well as noncontextual models correspond to non-negative solutions to a system of linear equations for the joint probability distributions of all observables.
	We demonstrate that nonclassical correlations correspond to a particular solution to this system, which may have negative values even if local realism is not violated.
	A modification of Bell inequalities enables us to test such correlations.
	At the same time, our approach leads to a formulation of Bell inequalities applicable also to continuous variables.
	The results are illustrated with two-mode squeezed vacuum states and with hybrid entangled states (Schr\"odinger-Cat states), one mode being analyzed by balanced and the other one by unbalanced homodyne detection.
\end{abstract}
	
	\maketitle


\section{Introduction}

    
    Unusual properties of correlations between spatially separated parts of quantum systems are of great interest since the pioneering work by Einstein, Podolsky, and Rosen \cite{einstein35}.
    Their ideas had been formalized by Bell \cite{bell64} by introducing the principles of local realism, which naturally describe classical correlations.
    Quantum states violating these principles exhibit the phenomenon of Bell nonlocality \cite{Brunner2014}, representing the strongest type of quantum correlations \cite{Wiseman2007,Jones2007,Adesso2016}.
    This means that the violation of local realism implies the presence of other correlations but not necessarily the other way around. 
    Lower levels at the hierarchical structure of these correlations are represented successively by quantum steering \cite{Wiseman2007} and quantum entanglement \cite{horodecki09,Werner1989}.  
    
  	Typically one characterizes quantum states by their principal abilities to manifest correlations.
    A more specific approach consists in characterization of quantum states together with particular measurements conducted independently by two parties, Alice and Bob. 
 	In this case we deal with the probability distributions $\mathcal{P}(A,B|a,b)$, also referred to as the behavior \cite{Tsirelson93,Brunner2014}, to get the values $A$ and $B$ given the measurement-device settings $a$ and $b$ for Alice and Bob, respectively.
 	It is also important that the device settings are chosen independently and randomly.
	
	
	The core idea of local realism consists in the assumption that system states can be characterized by non-negative probability distributions $\rho(\omega)\geq 0$ for a set of variables $\omega\in\Omega$.
	In turn, the considered measurement devices are characterized by the response functions $\mathrm{F}_{\mathrm{A}}(A|a;\omega)\geq 0$ and $\mathrm{F}_{\mathrm{B}}(B|b;\omega)\geq 0$ being the probabilities to get the values $A$ and $B$ given the variables $\omega$ and the settings $a$ and $b$, respectively.
	This yields the expression
        \begin{align}\label{Eq:LocalRealism}
            \mathcal{P}(A,B|a,b) = \int_{\Omega} \D \omega  \rho(\omega) \mathrm{F}_{\mathrm{A}}(A|a;\omega) \mathrm{F}_{\mathrm{B}}(B|b;\omega)
        \end{align}
	for the behavior.
	The nonexistence of non-negative constituents at the right-hand side of Eq.~(\ref{Eq:LocalRealism}) for a given behavior implies that the corresponding quantum state and measurements cannot be explained within the framework of local realism. 
 	Equivalently, local realism implies that the elements of the behavior can be considered as marginals of a non-negative joint probability distribution for all observables (JPDAO) \cite{wigner1970,fine82,kaszlikowski2000,abramsky2011}. 
 	If such a distribution exists, then the behavior can be explained by a local realistic model.
 	As it has been discussed in Ref.~\cite{abramsky2011}, such an interpretation is closely related to the concept of quantum contextuality \cite{Spekkens2008,Cabello2021}.
 	The concept of JPDAO has also been discussed in the context of classical phase-space distributions \cite{Matzkin2008}.
 
 	In order to prove the existence of a local realistic model, it is sufficient to find a non-negative \mbox{JPDAO} corresponding to the given behavior.
 	The nonexistence of local realistic models means that non-negative JPDAOs do not exist.
 	For a given behavior, this fact can be proven via solving the dual form \cite{Brunner2014} of a linear-programming problem leading to Bell inequalities \cite{bell64,CHSH}.
 	Their violation implies the presence of strong quantum correlations corresponding to Bell nonlocality.   
 	Originally, these inequalities have been presented for dichotomic observables with two settings for Alice and Bob.
 	Further, they have been generalized to different configurations; see, e.g., the corresponding list in Ref.~\cite{Brunner2014}.  	
 	        
	
	A different type of quantum correlations in quantum optics has been discussed in Refs.~\cite{Agudelo2013,shahandeh17b}.
	They are based on the fact that, similar to the single-mode scenario, any two-mode quantum state described by the density operator $\hat{\rho}$ can be expanded by coherent states $\ket{\alpha_{\mathrm{A}}}$ and $\ket{\alpha_{\mathrm{B}}}$,
	    \begin{align}\label{Eq:PFuncExp}
            \hat{\rho} = \int_{\mathbb{C}^2}\!\!\! \D^2\! \alpha_{\mathrm{A}} \D^2\! \alpha_{\mathrm{B}}
            P(\alpha_{\mathrm{A}}, \alpha_{\mathrm{B}})
            \ket{\alpha_{\mathrm{A}}} \bra{\alpha_{\mathrm{A}}} \otimes \ket{\alpha_{\mathrm{B}}} \bra{\alpha_{\mathrm{B}}},
        \end{align}
	where $P(\alpha_{\mathrm{A}}, \alpha_{\mathrm{B}})$ is the Glauber-Sudarshan $P$ function \cite{glauber63c,sudarshan63}.
	If $P(\alpha_{\mathrm{A}}, \alpha_{\mathrm{B}})\geq 0$, then the corresponding state can be simulated by a statistical mixture of coherent states.
	Otherwise, the state is considered as nonclassical \cite{titulaer65,reid1986,Hillery1987,Lee1991, mandel86,mandel_book,vogel_book,agarwal_book,sperling2018a,sperling2018b,sperling2020,vogel00,richter02,Asboth2005,vogel08,kiesel10,rivas2009,Miranowicz2015a,Miranowicz2015b,park2015a,park2015b,bohmann2020,bohmann2020b}. 
	A particular interest is attracted to cases for which the reduced local states for Alice and Bob are classical but the whole two-mode state is nonclassical.
	In this case, the impossibility of simulations with coherent states is caused solely by correlations between two parts of the quantum state.
	As it has been discussed in Refs.~\cite{Agudelo2013,shahandeh17b,Koehnke2021}, such correlations may be inherent  even for separable states as well as for states with zero quantum discord \cite{ollivier01,henderson01,ferraro12}. 
	

	In this paper we show that nonclassical correlations of radiation related to negativities of the Glauber-Sudarshan P function and to negativities of the more general $s$-parameterized Cahill-Glauber distributions \cite{cahill69,cahill69a} can be naturally linked to Bell nonlocality.
	Such correlations correspond to a particular form of the \mbox{JPDAO}, which may not be positive semidefinite even for states and measurements exhibiting local realism.
	We have shown that such correlations can be tested with modified Bell inequalities.
	Although we consider a nonlocal state, particular measurement procedures applied to it can still be described with local realistic models.
	We demonstrate that nonclassical correlations of radiation in such cases can still be tested with the derived inequalities.

	
	The rest of the paper is organized as follows.
	In Sec.~\ref{Sec:BellNonlProbl} we consider the general statement of the Bell-nonlocality problem expressed as a system of linear equations for the JPDAO. A	
	mathematical formulation for the problem of nonclassical correlations for radiation is considered in Sec.~\ref{Sec:PSNC}.
	In Sec.~\ref{Sec:Dual} we derive standard Bell inequalities and inequalities for nonclassical correlations of radiation as the dual forms of linear-programming problems.
	In Sec.~\ref{Sec:Hybrid} we consider hybrid measurements, wherein continuous (quadratures) and discrete (displaced numbers of photons) variables are measured at Alice's and Bob's side, respectively.
	A necessary and sufficient condition for the presence of Bell nonlocality in such a measurement scenario is considered in Sec.~\ref{Sec:CHSH}.
	Section~\ref{Sec:TMSVS} is devoted to an example of the two-mode squeezed vacuum states (TMSVS).  
	We demonstrate that for such states Bell nonlocality cannot be verified by conducting hybrid measurements, while nonclassical correlations introduced here in such a scenario still exist.
	An example of hybrid entangled states leading to violation of local realism with the scenario of a hybrid measurement scheme is considered in Sec.~\ref{Sec:CatState}.
	A summary and some concluding remarks are given in Sec.~\ref{Sec:Conclusions}.


    \section{Bell-nonlocality problem}
    \label{Sec:BellNonlProbl}
    
    As discussed in the Introduction, the concept of local realism requires the presence of at least two observers, Alice and Bob, conducting measurements at their sides with outcomes $A$ and $B$, respectively.
    Alice and Bob randomly change settings of their measurement devices, $a_i$ and $b_j$, correspondingly.
    Each setting implies measurement of a particular observable.
    For the sake of simplicity, we restrict our consideration to two settings at each side, also noting that generalizations to any number of settings and to their continuous set are straightforward.
    
\begin{figure}[ht!]
            \includegraphics[width=0.9\linewidth]{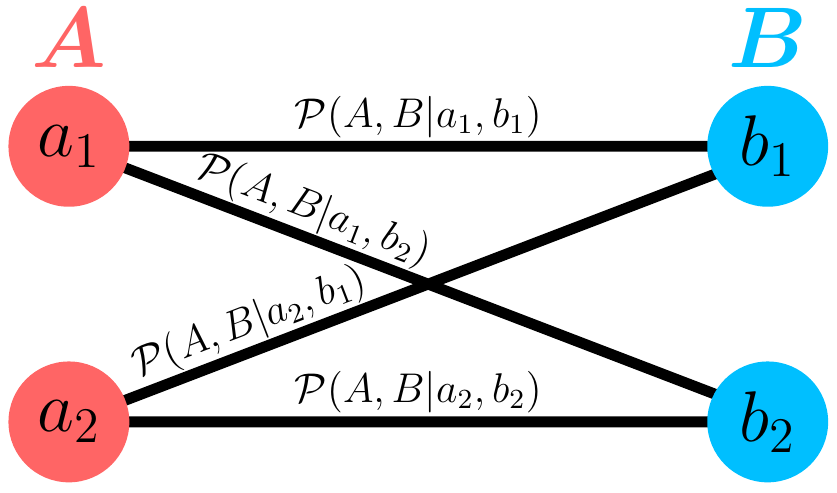}
            \caption{Scheme of the Bell-like experiment in the case of two settings at each side. 
            The measurement outcomes for Alice and Bob take values $A$ and $B$ with the settings $\{a_1,a_2\}$ and $\{b_1,b_2\}$, respectively.
            Four possible pairs of measurements are shown by lines.
            The corresponding conditional probabilities are designated as $\mathcal{P}(A,B|a_i,b_j)$.}
            \label{Fig:Bell}
        \end{figure}

    Let us consider the measurement scheme sketched in Fig.~\ref{Fig:Bell}.
    In the framework of local realism, the conditional probabilities $\mathcal{P}(A,B|a,b)$, cf. Eq.~(\ref{Eq:LocalRealism}), are considered as marginals of a non-negative JPDAO $\mathcal{W}(A_1,A_2,B_1,B_2)\equiv\mathcal{W}(\mathbf{A},\mathbf{B})$, which is a function of the sets of values $\mathbf{A}=\{A_1,A_2\}$ and $\mathbf{B}=\{B_1,B_2\}$ given the settings $\{a_1,a_2\}$ and $\{b_1,b_2\}$, respectively.
    This can be formalized as
        \begin{align}\label{Eq:FineLRM}
            \mathcal{P}(A,B|a_i,b_j)=\sum\limits_{\mathbf{A},\mathbf{B}}\mathcal{W}(A_1,A_2,B_1,B_2)\delta_{A,A_i}\delta_{B,B_j}, 
        \end{align}
    where $\delta_{A,A_i}$ and $\delta_{B,B_j}$ are Kronecker deltas.
    Equation (\ref{Eq:FineLRM}) can be considered as a system of linear algebraic equations for the JPDAO.
    If at least one observable takes continuous values, the corresponding sum is replaced with the integral and the Kronecker delta is replaced with the Dirac delta function.
    In that scenario, Eq.~(\ref{Eq:FineLRM}) becomes a system of linear integral equations. 
    The behavior $\mathcal{P}(A,B|a_i,b_j)$ can be explained within the framework of a local realistic model iff this system has at least one  non-negative solution $\mathcal{W}(A_1,A_2,B_1,B_2)\geq0$.   
   
    Equation~(\ref{Eq:FineLRM}) is a typical example of the inhomogeneous systems of linear equations.  
    Its general solution can be given by, cf. Ref.~\cite{strang_book_1988},
        \begin{align}\label{Eq:GenSolution}
           \mathcal{W}(A_1,&A_2,B_1,B_2)=\\
           &\mathcal{W}_\mathrm{h}(A_1,A_2,B_1,B_2)+\mathcal{W}_\mathrm{p}(A_1,A_2,B_1,B_2).\nonumber
        \end{align}
    Here $\mathcal{W}_\mathrm{h}(A_{1},A_{2},B_{1},B_{2})$ is the general solution to the corresponding homogeneous system, when the left-hand side of Eq.~(\ref{Eq:FineLRM}) is zero.
    The function $\mathcal{W}_\mathrm{p}(A_{1},A_{2},B_{1},B_{2})$ is a particular solution to the original inhomogeneous system.
    The function $\mathcal{W}_\mathrm{h}(A_1,A_2,B_1,B_2)$ depends also on arbitrary constants or, in the case of continuous variables, on arbitrary functions.
    If the local realistic model corresponding to the given behavior $\mathcal{P}(A,B|a_i,b_j)$ exists, then these constants (functions) can be chosen such that the function $\mathcal{W}(A_1,A_2,B_1,B_2)$ is non-negative.
    A particular solution $\mathcal{W}_\mathrm{p}(A_1,A_2,B_1,B_2)$, however, may have non-negative values.
    In the next section we will consider an important special case of $\mathcal{W}_\mathrm{p}(A_1,A_2,B_1,B_2)$ corresponding to phase-space nonclassical correlations.


\section{Nonclassical correlations of radiation}
\label{Sec:PSNC}
    
    In order to analyze nonclassical correlations of radiation \cite{Agudelo2013,shahandeh17b}, we remind ourselves that the behavior $\mathcal{P}(A,B|a,b)$ can be obtained from Born's rule \cite{born26} as
    	\begin{align}\label{Eq:BornRule}
    		\mathcal{P}(A,B|a,b)=\Tr\left[\hat{\rho}\,\hat{\Pi}_\mathrm{A}(A|a)\hat{\Pi}_\mathrm{B}(B|b)\right],
    	\end{align}	 
    where $\hat{\rho}$ is the density operator characterizing the quantum state of the light modes, and $\hat{\Pi}_\mathrm{A}(A|a)$ and $\hat{\Pi}_\mathrm{B}(B|b)$ are the positive operator-valued measures (POVM) \cite{nielsen10} characterizing measurements at Alice's and Bob's sides, respectively.
    Equation (\ref{Eq:BornRule}) can be rewritten in the phase-space representation, cf. Refs.~\cite{cahill69,cahill69a}:
		\begin{align}\label{Eq:PSLRM}
		    \mathcal{P}(A,B|a,b) 
		    &= \int_{\mathbb{C}^2} \D^4 \boldsymbol{\alpha} P(\alpha_\mathrm{A},\alpha_\mathrm{B};s_\mathrm{A},s_\mathrm{B})\nonumber\\
		    &\times\Pi_{\mathrm{A}}(A|a;\alpha_\mathrm{A};-s_\mathrm{A}) \Pi_{\mathrm{B}}(B|b;\alpha_\mathrm{B};-s_\mathrm{B}).
    	\end{align}
    Here $\D^4 \boldsymbol{\alpha}=\D^2\alpha_{\mathrm{A}}\D^2\alpha_{\mathrm{B}}$, $P(\alpha_\mathrm{A},\alpha_\mathrm{B};s_\mathrm{A},s_\mathrm{B})$ is the Cahill-Glauber $s$-parameterized quasiprobability distribution, and $\Pi_{\mathrm{A}}(A|a;\alpha_\mathrm{A};-s_\mathrm{A})$, $\Pi_{\mathrm{B}}(B|b;\alpha_\mathrm{B};-s_\mathrm{B})$ are $(-s)$-parameterized symbols of the POVM; see also Appendix~\ref{App:PSF} for details.
    The parameter $s\in[-1,1]$ describes the operator ordering. For example, the cases of $s=-1$, $s=0$, and $s=1$ correspond to antinormal, symmetric, and normal ordering, respectively.
    	
    Let us consider the case of $s_{\mathrm{A}}=s_{\mathrm{B}}=1$.
    This implies that $P(\alpha_\mathrm{A},\alpha_\mathrm{B};s_\mathrm{A},s_\mathrm{B})=P(\alpha_\mathrm{A},\alpha_\mathrm{B})$ is the Glauber-Sudarshan P function.
    The POVM symbols in this case are Q symbols,
    	\begin{align}
    		&\Pi_{\mathrm{A}}(A|a;\alpha_\mathrm{A})=\bra{\alpha_\mathrm{A}}\hat{\Pi}_\mathrm{A}(A|a)\ket{\alpha_\mathrm{A}}\geq0,\label{Eq:POVM_QA}\\
    		&\Pi_{\mathrm{B}}(B|b;\alpha_\mathrm{B})=\bra{\alpha_\mathrm{B}}\hat{\Pi}_\mathrm{B}(B|b)\ket{\alpha_\mathrm{B}}\geq0.\label{Eq:POVM_QB}
    	\end{align}
    They can be interpreted as probability distributions to get values $A/B$ given the settings $a/b$ and the coherent states $\ket{\alpha_\mathrm{A}}/\ket{\alpha_\mathrm{B}}$ at Alice's/Bob's side, respectively.
    If $P(\alpha_\mathrm{A},\alpha_\mathrm{B})\geq0$, then Eq.~(\ref{Eq:PSLRM}) is a particular case of Eq.~(\ref{Eq:LocalRealism}), i.e., it represents a local realistic model.
    This means that the measurement outcomes of such a state can be simulated by statistical mixture of coherent states.
    
    As it has been discussed in Ref.~\cite{Semenov2021} for single-party scenarios, simulations of measurement outcomes with coherent states may be possible even if $P(\alpha_\mathrm{A},\alpha_\mathrm{B})\ngeq0$.
    In order to demonstrate that, consider Eq.~(\ref{Eq:PSLRM}) as a system of integral equations with respect to the unknown function $P(\alpha_\mathrm{A},\alpha_\mathrm{B})$ given  $\Pi_{\mathrm{B}}(B|b;\alpha_\mathrm{B})$, $\Pi_{\mathrm{A}}(A|a;\alpha_\mathrm{A})$, and $\mathcal{P}(A,B|a,b)$.
    If the measurement is not informationally complete \cite{busch91,DAriano2004,prugovecki77,Schroeck_book,renes2004}, then the Glauber-Sudarshan P function is not the single solution to this system of equations.
    Non-negative functions can be among other solutions, and thus the measurements can be simulated with coherent states.
    
    Negativities of the function $P(\alpha_\mathrm{A},\alpha_\mathrm{B})$ can be associated with both nonclassical correlations and nonclassicality of local states at Alice's/Bob's sides.
    Similar to considerations in Refs.~\cite{Agudelo2013,shahandeh17b}, the uncorrelated density operator,
    	\begin{align}\label{Eq:UncorrStates}
    		\hat{\rho}_\mathrm{u}=\Tr_\mathrm{B}\hat{\rho}\otimes\Tr_\mathrm{A}\hat{\rho}, 
    	\end{align}
    can be assigned to the density operator $\hat{\rho}$.	
    Here $\Tr_{B/A}$ is the partial trace of the mode $B/A$, respectively.
    If simulations with coherent states are impossible for the state $\hat{\rho}$ but possible for the state $\hat{\rho}_\mathrm{u}$, nonclassical phenomena are solely related to correlations between modes \cite{Agudelo2013,shahandeh17b}.  
    
    One can also consider a more general type of correlations.
    For each type of measurements, there exist threshold values of $s=s_\mathrm{th}$ such that $\Pi_{\mathrm{A}}(A|a;\alpha_\mathrm{A};-s_\mathrm{th})\geq 0$ and $\Pi_{\mathrm{A}}(A|a;\alpha_\mathrm{A};-s)\ngeq 0$ for $s<s_\mathrm{th}$.
    If Eq.~(\ref{Eq:PSLRM}) has non-negative solutions $P(\alpha_\mathrm{A},\alpha_\mathrm{B};s_\mathrm{A},s_\mathrm{B})$ for $s_\mathrm{A}$ and $s_\mathrm{B}$ equal to or exceeding the corresponding threshold values, this equation determines a local realistic model.
     If such a model does not exist for the threshold values, it does not exist for all values of  $s_\mathrm{A}$ and $s_\mathrm{B}$.
     This means that phase-space classical simulations for such states are impossible \cite{Semenov2021}.
       
    Phase-space representation can be used for finding a particular solution to Eq.~(\ref{Eq:FineLRM}).
    Particularly, it can be used in Eq.~(\ref{Eq:GenSolution}) to find the general solution for the Bell-nonlocality problem.
    In order to demonstrate it, we consider a family of functions
         \begin{align}\label{Eq:DevDep}
            \mathcal{W}_\mathrm{p}(A_{1},A_{2},B_{1},B_{2})=
            \int_{\mathbb{C}^2}\D^4\boldsymbol{\alpha}
            P(\alpha_A,\alpha_B;s_A,s_B)\nonumber\\
            \times\Pi_{\mathrm{A}}(A_1|a_1;\alpha_\mathrm{A};-s_\mathrm{A})
            \Pi_{\mathrm{A}}(A_2|a_2;\alpha_\mathrm{A};-s_\mathrm{A})\nonumber\\  
            \times
            \Pi_{\mathrm{B}}(B_1|b_1;\alpha_\mathrm{B};-s_\mathrm{B})
            \Pi_{\mathrm{B}}(B_2|b_2;\alpha_\mathrm{B};-s_\mathrm{B}),
        \end{align}
    where $P(\alpha_A,\alpha_B;s_A,s_B)$ is a solution to Eq.~(\ref{Eq:PSLRM}) including phase-space quasiprobability distributions.
    Evidently, the functions $\mathcal{W}_\mathrm{p}(A_{1},A_{2},B_{1},B_{2})$  represents a family of particular solutions to Eq.~(\ref{Eq:FineLRM}).
    This fact can be proven by using the normalization conditions for the POVM,
        \begin{align}
            \sum_{A}\Pi_{\mathrm{A}}(A|a;\alpha_\mathrm{A};-s_\mathrm{A})=\sum_{B}\Pi_{\mathrm{B}}(B|b;\alpha_\mathrm{B};-s_\mathrm{A})=1.
        \end{align}
    Possibilities of phase-space classical simulations, including simulations with coherent states, imply that $\mathcal{W}_\mathrm{p}(A_{1},A_{2},B_{1},B_{2})\geq0$.
    The quantum state and the measurements, i.e., the behavior $\mathcal{P}(A,B|a,b)$, cannot be simulated with phase-space functions if $\mathcal{W}_\mathrm{p}(A_{1},A_{2},B_{1},B_{2})\ngeq0$.
    In particular, this is related to nonclassical correlations of radiation.  
    
	Since the function (\ref{Eq:DevDep}) is a particular solution to Eq.~(\ref{Eq:FineLRM}), Bell nonlocality always implies the presence of nonclassical correlations of radiation.
	The inverse statement is, in general,  untrue.
	For example, nonclassical correlations of radiation can be found in separable states, cf.  Refs.~\cite{Agudelo2013,shahandeh17b}.
	Local realistic models for such states exist for any measurement. 

	For another scenario, we refer to nonlocal states.
	However, particular measurements applied to them can still be described with a local realistic model.
	This corresponds to local realistic behaviors $\mathcal{P}(A,B|a,b)$ for Bell nonlocal states.
	In the following, we will show that such behaviors can also be nonclassical, including the case of nonclassical correlations of radiation.
	
\begin{figure}[ht!]
            \includegraphics[width=0.9\linewidth]{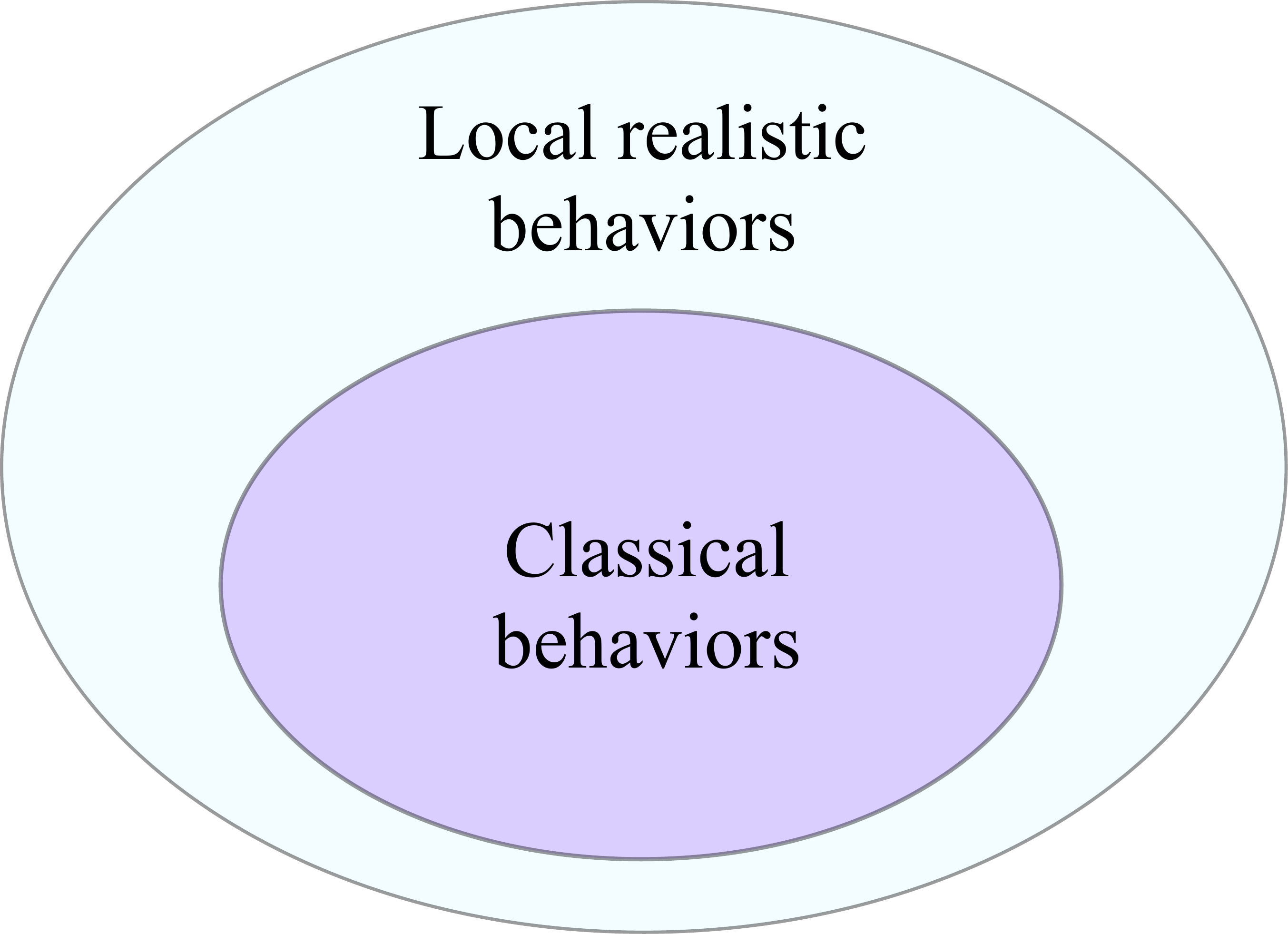}
            \caption{\label{Fig:Venn} The Venn diagram demonstrate relations between the sets of local realistic and classical behaviors.}
        \end{figure}

\begin{figure}[ht!]
            \includegraphics[width=0.9\linewidth]{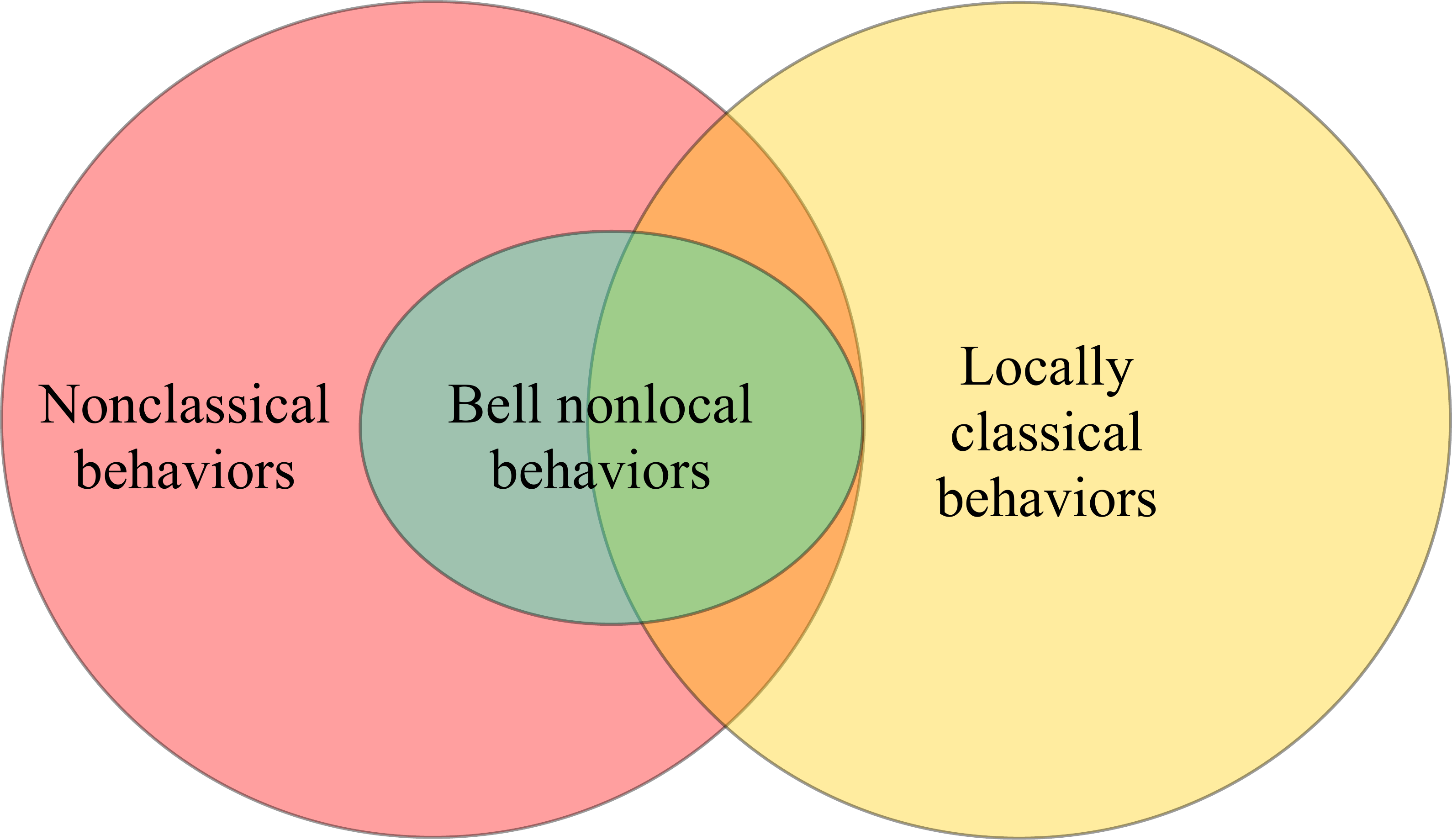}
            \caption{\label{Fig:VennLCB} The Venn diagram demonstrates relations between the sets of Bell nonlocal, nonclassical, and locally classical behaviors.
            The behaviors demonstrating nonclassical correlations and no local nonclassicality belong to the intersection of sets of locally classical and nonlocal behaviors.}
        \end{figure}

	In general, we consider nonclassicality and nonlocality of behaviors $\mathcal{P}(A,B|a,b)$.
	The classical behaviors, for which non-negative solutions to Eq.~(\ref{Eq:PSLRM}) exist, are always local realistic, i.e., they can be explained by Eq.~(\ref{Eq:LocalRealism}).
	The inverse statement is not true.
	The corresponding relations between these sets are shown in Fig.~\ref{Fig:Venn}. 
	A more complicated relation occurs if we consider also locally classical behaviors.
	In order to define this notion, we first assign the factorized behavior
	    \begin{align}\label{Eq:FactBehavior}
	        \mathcal{P}_\mathrm{u}(A,B|a,b)=\mathcal{P}(A|a)\mathcal{P}(B|b)
	    \end{align}
	to each behavior $\mathcal{P}(A,B|a,b)$, where 
	    \begin{align}
                \mathcal{P}(A|a)=\sum\limits_{B}\mathcal{P}(A,B|a,b)\label{Eq:MargA}\\
                \mathcal{P}(B|b)=\sum\limits_{A}\mathcal{P}(A,B|a,b)\label{Eq:MargB}
        \end{align}
    are marginal probability distributions.
    The behavior $\mathcal{P}(A,B|a,b)$ is locally classical if Eq.~(\ref{Eq:PSLRM}) with the corresponding $\mathcal{P}_\mathrm{u}(A,B|a,b)$ has a non-negative solution.
    The relation between the sets of Bell nonlocal, nonclassical, and locally classical behaviors is shown in Fig.~\ref{Fig:VennLCB}.
    Intersection of the sets of locally classical and nonclassical behaviors corresponds to behaviors demonstrating nonclassical correlations of radiation \cite{Agudelo2013,Koehnke2021}.
    	


\section{Dual form of quantum-correlation problems}
\label{Sec:Dual}

    If we aim to explain the given behavior $\mathcal{P}(A,B|a,b)$ with a local realistic model, we have to find at least one non-negative solution to Eq.~(\ref{Eq:FineLRM}).
    In order to prove that such models do not exist, we can employ the dual form of this problem \cite{Brunner2014}, expressed in the form of the Bell inequalities directly following from the hyperplane separation theorem \cite{boyd_book}.
    Their violations imply the nonexistence of local realistic models related to the considered behavior.
    
    A similar situation takes place for nonclassical correlations of radiation.
    When we aim to prove their nonexistence, it suffices to show that one of the particular phase-space solutions (\ref{Eq:DevDep}) to Eq.~(\ref{Eq:FineLRM}) is non-negative, or one can find a non-negative solution $P(\alpha_\mathrm{A},\alpha_\mathrm{B};s_\mathrm{A},s_\mathrm{B})$ to Eq.~(\ref{Eq:PSLRM}).
    The presence of such correlations requires one to prove the fact that all such solutions have negativities, which can be a complicated task, also in the case of real experimental data.
    Similar to the case of Bell nonlocality, we can apply here the hyperplane separation theorem.
    This leads to an analog of Bell inequalities.
    Their violations imply the presence of nonclassical correlations of radiation.
    
    \subsection{General formulation}
    
    The direct forms of the problems of Bell nonlocality and nonclassical correlations of radiation can be both formulated as the linear equation,
        \begin{align}\label{Eq:DirectProblem}
            \bm{\mathcal{P}}=\mathcal{M}\textbf{W}.
        \end{align}
    Here $\bm{\mathcal{P}}$ is the vector (also from a function space) composed from all values of the behavior.
    The vector $\textbf{W}$ is composed from all values of the JPDAO in the case of the Bell-nonlocality problem, cf. Eq.~(\ref{Eq:FineLRM}).
    For the problem of nonclassical correlations of radiation this vector is given by the set of all values of $P(\alpha_\mathrm{A},\alpha_\mathrm{B};s_\mathrm{A},s_\mathrm{B})$, cf. Eq.~(\ref{Eq:PSLRM}).
    With the symbol $\mathcal{M}$, we denote the mapping from $\textbf{W}$ to  $\bm{\mathcal{P}}$.
    This mapping has a specific form for each problem.
    
    As it follows from Eq.~(\ref{Eq:DirectProblem}), all $\bm{\mathcal{P}}$ corresponding to non-negative $\textbf{W}$ form a convex set.
    Hence, the hyperplane separation theorem resulting in Bell-like inequalities can be applied here.
    Let us consider an arbitrary vector $\mathbf{f}$.
    If all components of $\mathbf{W}$ are non-negative, then $\mathbf{f}^\mathrm{T}\cdot\mathbf{W}\leq\sup\mathbf{f}(\mathbf{1}\cdot\mathbf{W})$, where all components of the vector (values of the function) $\mathbf{1}$ are unity.\footnote{Here $\sup\mathbf{f}$ means supremum of $\mathbf{f}$ as a function of a discrete, continuous, or hybrid set of variables \cite{Rudin_book}.
    For example, in the case of discrete finite sets of variables, it corresponds to the maximal component of the vector.}
    In fact, this means that the upper bound of the left-hand side of this inequality is obtained by replacing each component of the vector (each value of the function) $\mathbf{f}$ with its supremum. 
    Normalization of $\mathbf{W}$ implies $(\mathbf{1}\cdot\mathbf{W})=1$.
    This yields $\mathbf{f}^\mathrm{T}\cdot\mathbf{W}\leq\sup\mathbf{f}$.
    Taking the scalar product for both sides of Eq.~(\ref{Eq:DirectProblem}) with an arbitrary vector $\boldsymbol{\lambda}$ and considering $\mathbf{f}^\mathrm{T}=\boldsymbol{\lambda}^\mathrm{T}\mathcal{M}$ we arrive at the inequalities
        \begin{align}\label{Eq:GenIneq}
           \boldsymbol{\lambda}^\mathrm{T}\cdot\bm{\mathcal{P}}\leq \sup\boldsymbol{\lambda}^\mathrm{T}\mathcal{M}.
        \end{align}
    If there exist a $\boldsymbol{\lambda}$ such that these inequalities are violated, then the given behavior manifests Bell nonlocality or nonclassicality of radiation, depending on the used mapping $\mathcal{M}$.
    It can also be shown that the inverse statement is true: if a non-negative $\mathbf{W}$ does not exist, then there exist $\boldsymbol{\lambda}$ such that these inequalities are violated, cf. Refs.~\cite{Farkas1902,Brunner2014}.

    \subsection{Bell inequalities}
        
        In this section we recall the standard Bell inequalities used for testing local realism \cite{bell64, Brunner2014}.
        In this case, the matrix (integral kernel) $\mathcal{M}$ is given by  $\delta_{A,A_i}\delta_{B,B_j}$  in Eq.~(\ref{Eq:FineLRM}). 
        Applying this particular form in Eq.~(\ref{Eq:GenIneq}), one gets   
            \begin{align}\label{Eq:BellIneq}
                \sum_{i,j} E(\lambda|a_i,b_j) \leq \sup_{\mathbf{A}, \mathbf{B}} \sum_{i,j} \lambda(A_i,B_j,a_i,b_j),
            \end{align}
        where 
            \begin{align}\label{Eq:ExpectedLambda}
                E(\lambda|a_i,b_j) = \sum_{A,B} \lambda(A,B,a_i,b_j) \mathcal{P}(A,B|a_i,b_j)
            \end{align}
        is the expected value of $\lambda(A,B,a_i,b_j)$ over the random variables $A$ and $B$ given the settings $a_i, b_j$.  
        The supremum at the right-hand side of this equation is evaluated over all components of the vectors $\mathbf{A}=\{A_1,A_2\}$ and $\mathbf{B}=\{B_1,B_2\}$.
        In the case that at least one observable takes continuous values, the corresponding sum in Eq.~(\ref{Eq:ExpectedLambda}) is replaced by integration.
        
        Inequality~(\ref{Eq:BellIneq}) represents the most general form of linear Bell inequalities.
        If there exists $\lambda(A,B,a_i,b_j)$ such that these inequalities are violated, then the behavior $\mathcal{P}(A,B|a_i,b_j)$ exhibits Bell nonlocality.
        Particular choices of $\lambda$ result in Bell inequalities for different measurement configurations.
        For example, a special choice of $\lambda$ in the case of dichotomic variables leads to the well-known Clauser-Horn-Shimony-Holt (CHSH) inequalities \cite{CHSH}, see Appendix~\ref{App:CHSH}.
        They represent a necessary and sufficient condition for existence of a local realistic model in the appropriate measurement configuration.
        In the most general case, finding such complete and tight sets of $\lambda$, cf. Ref.~\cite{Brunner2014}, is a computationally hard problem.
        However, for the given behavior $\mathcal{P}(A,B|a_i,b_j)$, it can be reformulated: we may look for a special $\lambda$ violating inequality (\ref{Eq:BellIneq}).
        
        The expected values $E(\lambda|a_i,b_j)$ at the left-hand side of Bell inequalities (\ref{Eq:BellIneq}) can be directly obtained from experimental data.
        Typically, such data are given by $M$ pairs of measured observables, $(A^{(k)},B^{(k)})$, where $k=1\ldots M$, for each setting $a_i,b_j$.
        Thus the expected values can be estimated as
            \begin{align}
                E(\lambda|a_i,b_j)\approx \frac{1}{M}\sum\limits_{k=1}^{M}\lambda(A^{(k)},B^{(k)},a_i,b_j). \label{Eq:ExpEstimation}
            \end{align}
        Therefore, while the right-hand side of Bell inequalities (\ref{Eq:BellIneq}) is fixed by a particular choice of $\lambda$, its left-hand side is given by the measurement outcomes.

    \subsection{Bell-like inequalities for nonclassical correlations of radiation}
        
        In this section we generalize the approach of Bell-like inequalities for phase-space classical simulations \cite{Semenov2021} to the case of nonclassical correlations of radiation.
        For this purpose we will use the integral kernel $\mathcal{M}$ in the form of $\Pi_{\mathrm{A}}(A|a;\alpha_\mathrm{A};-s_\mathrm{A}) \Pi_{\mathrm{B}}(B|b;\alpha_\mathrm{B};-s_\mathrm{B})$, and the role of the vector $\mathbf{W}$ from a function space is played by the functions $P(\alpha_\mathrm{A},\alpha_\mathrm{B};s_\mathrm{A},s_\mathrm{B})$.  
        Inequality (\ref{Eq:GenIneq}), following from the hyperplane separation theorem, in this case is reduced to the form
            \begin{align}\label{Eq:PSBell}
                \sum_{i,j}E(\lambda|a_i,b_j) \leq \sup_{\boldsymbol{\alpha}} \sum_{i,j} E(\lambda|a_i,b_j;\boldsymbol{\alpha};s_\mathrm{A},s_\mathrm{B}), 
            \end{align}
        where the left-hand side is still given by Eq.~(\ref{Eq:ExpectedLambda}).
        The sum components at the right-hand side of this inequality are given by the expression
            \begin{align}\label{Eq:CoherentMean}
                E(\lambda&|a_i,b_j;\boldsymbol{\alpha},s_\mathrm{A},s_\mathrm{B})  = \sum_{A,B} \lambda(A,B,a_i,b_j) \nonumber \\  &\times \Pi_{\mathrm{A}}(A|a_i;\alpha_{\mathrm{A}};-s_{\mathrm{A}}) \Pi_{\mathrm{B}}(B|b_j;\alpha_{\mathrm{B}};-s_{\mathrm{B}}).
            \end{align}   
        In both equations we use the vector $\boldsymbol{\alpha}=(\alpha_{\mathrm{A}},\alpha_{\mathrm{B}})$.
        
        Inequalities (\ref{Eq:PSBell}) have a form which is similar to standard Bell inequalities (\ref{Eq:BellIneq}) but with a modified right-hand side.
        Their violations imply the impossibilities of phase-space classical simulations for the given behavior $\mathcal{P}(A,B|a_i,b_j)$.
        Among others, this means that the particular solution (\ref{Eq:DevDep}) for the Bell-nonlocality problem (\ref{Eq:FineLRM}) is not positive semidefinite for any $P(\alpha_A,\alpha_B;s_A,s_B)$, being a solution to Eq.~(\ref{Eq:PSLRM}). 
        This can be caused by both nonclassicality of local states and by the presence of nonclassical correlations of radiation.
        For the latter case, such inequalities are not violated with the factorized behavior $\mathcal{P}_\mathrm{u}(A,B|a,b)$ given by Eq.~(\ref{Eq:FactBehavior}).
        In particular, this implies that replacing $E(\lambda|a_i,b_j)$ with its uncorrelated counterpart,
            \begin{align}\label{Eq:ExpectedLambdaSep}
                E_\mathrm{u}(\lambda|a_i,b_j) = \sum_{A,B} \lambda(A,B,a_i,b_j) \mathcal{P}_\mathrm{u}(A,B|a_i,b_j),
            \end{align}
        does not result in violation of inequalities (\ref{Eq:PSBell}).
        Fulfillment of inequalities (\ref{Eq:PSBell}) with $E_\mathrm{u}(\lambda|a_i,b_j)$ for a given $\lambda$ is only a necessary condition for the fact that their violation with $E(\lambda|a_i,b_j)$ is caused solely by nonclassical correlations.
        A sufficient condition of this fact is given by the existence of a non-negative solution of Eq.~(\ref{Eq:PSLRM}) with the left-hand side replaced by $\mathcal{P}_\mathrm{u}(A,B|a,b)$.    
        
        Let us consider the special case of $s_\mathrm{A}=s_\mathrm{B}=1$.
        As it has been already discussed, this scenario corresponds to the possibility of simulating correlations with coherent states.
        If inequality (\ref{Eq:PSBell}) is violated, then the measurement outcomes cannot be simulated by a statistical mixture of coherent states.
        It is worth noting that in this case the right-hand side of this inequality can be estimated from the experimental data as well.
        Taking into account Eqs.~(\ref{Eq:POVM_QA}) and (\ref{Eq:POVM_QB}), we conclude that $E(\lambda|a_i,b_j;\boldsymbol{\alpha};1,1)$, cf. Eq. (\ref{Eq:CoherentMean}), is the expected value of the quantity $\lambda(A,B,a_i,b_j)$ given the coherent state $\ket{\alpha_\mathrm{A}}\otimes\ket{\alpha_\mathrm{B}}$.
        This implies that the values given by Eq.~(\ref{Eq:CoherentMean}) can be estimated in the same way as it is described by Eq. (\ref{Eq:ExpEstimation}) but with coherent states at the source.
        If we are interested in checking the inequalities for other values of $s_\mathrm{A},s_\mathrm{B}$, the value of $E(\lambda|a_i,b_j;\boldsymbol{\alpha},s_\mathrm{A},s_\mathrm{B})$ can be obtained from $E(\lambda|a_i,b_j;\boldsymbol{\alpha};1,1)$ with the standard deconvolution formula connecting different phase-space symbols \cite{cahill69,cahill69a}.
        
        As mentioned, inequalities (\ref{Eq:PSBell}) are a generalization of the inequalities derived in Ref.~\cite{Semenov2021}.
        In turn, the latter can be considered as a generalization of inequalities presented in Ref.~\cite{rivas2009}.
        They are also related to the CHSH-like inequalities for testing nonclassicality discussed in Refs.~\cite{park2015a,park2015b}.    


    \section{Hybrid measurement scheme}
	\label{Sec:Hybrid}

    In this paper we consider two archetypal measurement procedures: balanced \cite{yuen1980,yuen1983,schumaker1984,yurke1987,schleich_book,VogelReview,vogel_book} and unbalanced \cite{wallentowitz96,mancini1997} homodyne detection. 
    Balanced homodyne detection is described by the POVM symbols given by 
    	\begin{align}\label{Eq:POVM_BHD}
    		\Pi(x|\varphi;\alpha;-s)=\frac{1}{\sqrt{\pi s}}\exp\left\{-\frac{\left[x{-}\sqrt{2}\Re\left(\alpha e^{-i\varphi}\right)\right]^2}{s}\right\},
    	\end{align}
    where $s\in[0,1]$, $x\in\mathbb{R}$ is the field quadrature with the phase $\varphi$, playing the role of the observable and the measurement-device setting, respectively.
    The POVM symbol of the unbalanced homodyne detection reads
        \begin{align} \label{Eq:POVM_UHD}
		    \Pi(n|\gamma ;\alpha)=\left( 1-e^{-|\alpha -\gamma |^{2}}\right)
		    ^{n}e^{-|\alpha -\gamma |^{2}(1-n)}.
	    \end{align}	
    Here the ordering parameter may take only a single value $s=1$, the observable $n\in \{0,1\}$ corresponds to no-click and click events at the detector, and the local-oscillator amplitude $\gamma$ plays the role of the measurement-device setting. 
    
    Measurement outcomes of the balanced and unbalanced homodyne detection are given by continuous and discrete variables, respectively.
    In the schemes checking Bell nonlocality, one usually considers identical measurements from each side.
    Typically, it is given by discrete variables, as, for example, in the original Bell proposal \cite{bell64,CHSH}. 
    Continuous variables are commonly analyzed via discretization of the outcomes  \cite{nha2004,Garcia2004}.
    Another approach \cite{cavalcanti2007} applied inclusively to continuous variables, using inequalities of the probability theory for functions of all observables. 
    Bell inequalities (\ref{Eq:BellIneq}) considered in this paper can also be directly applied to continuous variables.

\begin{figure}[ht!]
	    	\includegraphics[width=0.9\linewidth]{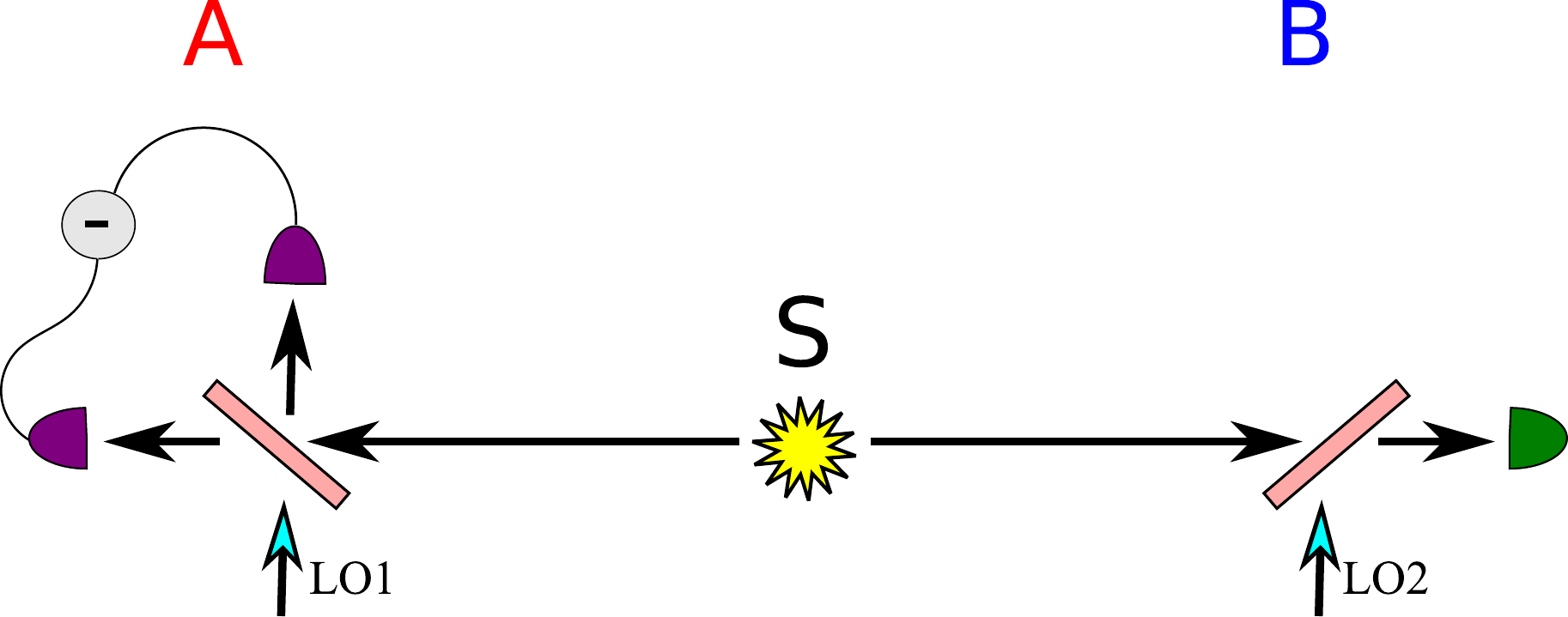}
	    	\caption{The scheme of hybrid measurements.
	    		Alice (A) and Bob (B) conduct balanced and unbalanced homodyne detection, respectively.
	    		Measurement-device settings are controlled by local oscillators LO1 and LO2. 
	    		The measurement outcomes for Alice and Bob take values $x\in\mathbb{R}$ and $n\in\{0,1\}$ with the settings $\varphi_i$ and $\gamma_j$, respectively, where $i,j=1,2$.
	    		}
	    	\label{Fig:HybrSch}
	    \end{figure}

    We concentrate on a hybrid measurement scheme, Fig.~\ref{Fig:HybrSch}, which uses balanced and unbalanced homodyne detection for Alice's and Bob's sides, respectively.
    This implies that Alice obtains continuous outcomes while Bob obtains discrete ones.
    The measurement-device settings $\varphi_i$ and $\gamma_j$ determined by the corresponding local oscillators take two discrete values, i.e., $i,j=1,2$.

    Specifying the measurement outcomes for Eq.~(\ref{Eq:FineLRM}) as ($A=x,a=\varphi$) and ($B=n,b=\gamma$), one gets
        \begin{align}\label{Eq:pMw_hom}
             &\mathcal{P}(x,n|\varphi_i,\gamma_j)=\\&\sum\limits_{n_1=0}^{1}\sum\limits_{n_{2}=0}^{1}\int_{\mathbb{R}}{\rm d}x_1\int_{\mathbb{R}}{\rm d}x_{2}
             \delta(x-x_i) \delta_{n,n_j}
             \mathcal{W}(x_1,x_{2},n_1,n_{2}).\nonumber
        \end{align}
  	The general solution to the homogeneous system in this case is given by
        \begin{align}
            \label{Eq:HomogeneousSolution}\mathcal{W}_\mathrm{h}(x_1,x_2,&n_1,n_2)=\\
            &(-1)^{n_1+n_2}C(x_1,x_2)+\delta_{n_1,0}\delta_{n_2,0}C_{00}(x_1,x_2)\nonumber\\
            +&\delta_{n_1,0}\delta_{n_2,1}C_{01}(x_1,x_2) +
            \delta_{n_1,1}\delta_{n_2,0}C_{10}(x_1,x_2).\nonumber
        \end{align}
    It depends on the arbitrary function $C(x_1,x_2)$ and on the functions $C_{00}(x_1,x_2)$, $C_{01}(x_1,x_2)$, $C_{10}(x_1,x_2)$ obeying the conditions
        \begin{align}
             \int\limits_{\mathbb{R}}{\rm d}x_1\, C_{ij}(x_1,x_2)=\int\limits_{\mathbb{R}}{\rm d}x_2\, C_{ij}(x_1,x_2)=0.
        \end{align}
   A particular solution to the inhomogeneous system can be found in the form of Eq.~(\ref{Eq:DevDep}).
   
\section{Generalized CHSH inequalities for the hybrid measurement scheme}
\label{Sec:CHSH}
    Here we apply the technique developed in previous sections and ideas of Ref.~\cite{Pironio2014} regarding generalized CHSH inequalities to derive a necessary and sufficient condition of locality for the hybrid measurement scheme.
    In particular, we will specify a set of test functions $\lambda^{(k)}(x,n|\varphi,\gamma)$, $k=1,2$ such that any nonlocal behavior violates at least one corresponding inequality (\ref{Eq:BellIneq}).
    These functions depend on the behavior $\mathcal{P}(x,n|\varphi_i, \gamma_j)$.
    Therefore, given the behavior, one can define the corresponding test functions and then check whether it is nonlocal.
    
    The considered test functions correspond to a discretization of the outcomes $x_i$ with nontrivial subsets.
    In order to describe them properly, we use the indicator function of a subset $X\in\mathbb{R}$,
        \begin{align}
            \mathbf{I}(x_i;X)=
            \left\{
            \begin{array}{lcr}
                1 & \textrm{if} & x_i\in X\\
                0 & \textrm{if} & x_i\notin X 
            \end{array}
            \right..
        \end{align}
    A necessary and sufficient condition for the locality of a given behavior $\mathcal{P}(x,n|\varphi_i, \gamma_j)$ can be formulated as the following statement.
    
    \begin{statement}
		The behavior $\mathcal{P}(x,n|\varphi_i, \gamma_j)$ for the hybrid measurement scheme is local iff it satisfies inequalities (\ref{Eq:BellIneq}) with the two test functions $\lambda^{(k)}(x,n|\varphi,\gamma)$, $k=1,2$, given by
			\begin{align}\label{Eq:HybridSchemeLambdas}
			   \lambda^{(k)}&(x,n|\varphi_i,\gamma_j)=\nonumber\\
			   &\left\{
			    \begin{array}{lcr}
			        -\mathbf{I}\left(x;X_1^{(k)}\right)\delta_{n, 1} & \textrm{if} & i=1,j=1\\
			        \mathbf{I}\left(x;X_1^{(k)}\right)\delta_{n, 0} & \textrm{if} & i=1,j=2\\
			        -\mathbf{I}\left(x;X_1^{(k)}\right)\delta_{n, 0} & \textrm{if} & i=2,j=1\\
			        1-\mathbf{I}\left(x;X_2^{(k)}\right)\delta_{n, 0} & \textrm{if} & i=2,j=2
			    \end{array}
			   \right.,
			\end{align}
		where
			\begin{align}
				\label{Eq:PartitionSets1}
				X_1^{(k)} = & \left\{ x \in \mathbb{R} | \mathcal{P}(x,0|\varphi_k,\gamma_1) + \mathcal{P}(x,0|\varphi_k,\gamma_2) \nonumber \right.\\
				& \left. - \mathcal{P}(x|\varphi_k) \geq 0 \right\}, \\
				X_2^{(k)} = & \left\{ x \in \mathbb{R} | \mathcal{P}(x,0|\varphi_l,\gamma_2) \geq \mathcal{P}(x,0|\varphi_l,\gamma_1) \right\}.
			\end{align}
		In these relations $l=1$ if $k=0$, $l=0$ if $k=1$, and
        \begin{align}\label{Eq:DistrX}
 	  	 \mathcal{P}(x|\varphi)=\sum\limits_{n=0}^1\mathcal{P}(x,n|\varphi,\gamma)
        \end{align}
       is the marginal distribution of $\mathcal{P}(x,n|\varphi,\gamma)$ for the quadrature $x$ given the phase $\varphi$.
	\end{statement}%
    
	\begin{proof}
	The necessity of this condition is obvious.
	Indeed, inequalities (\ref{Eq:BellIneq}) are fulfilled for any local behavior.
	In order to prove sufficiency we will show that if these inequalities with the test function (\ref{Eq:HybridSchemeLambdas}) are fulfilled, then it is always possible to construct a non-negative JPDAO.
		
	First, we note that the right-hand side of inequality (\ref{Eq:BellIneq}) with the test function (\ref{Eq:HybridSchemeLambdas}) is zero.
	This gives a possibility to rewrite the considered Bell inequalities in the form 
		  \begin{align}
            & \langle m \rangle_{\varphi_1} \leq \langle M \rangle_{\varphi_2}, \label{Eq:PironioCondition1}\\
            & \langle m \rangle_{\varphi_2} \leq \langle M \rangle_{\varphi_1} \label{Eq:PironioCondition2}.
          \end{align}
	Here we use the notations
		  \begin{align}
             & \langle m \rangle_{\varphi_1} = \int_{\mathbb{R}} \D x m(x,\varphi_1) , \\
             & \langle M \rangle_{\varphi_2} = \int_{\mathbb{R}} \D x M(x,\varphi_2),
          \end{align}
	where
		  \begin{align}
             m(x,\varphi) = & \max \{ \mathcal{P}(x,0|\varphi,\gamma_1) + \mathcal{P}(x,0|\varphi,\gamma_2) \nonumber \\
             & - \mathcal{P}(x|\varphi), 0 \}, \label{Eq:mExpression}\\
             M(x,\varphi) = & \min \{ \mathcal{P}(x,0|\varphi,\gamma_1), \mathcal{P}(x,0|\varphi,\gamma_2) \} \label{Eq:MExpression}
         \end{align}
    are non-negative functions of the quadrature $x$ and its phase $\varphi$ for the given settings $\gamma_1$ and $\gamma_2$.  
		
    Next we consider a class of the possible JPDAO for which the condition
      \begin{align}
          \mathcal{W}(x_i|n_1,n_2,x_j) = \mathcal{W}(x_i|n_1,n_2)
      \end{align}
    holds true, where $i,j = 1,2$ and $i \neq j$.
    This enables us to present $\mathcal{W}(x_1,x_2,n_1,n_2)$ as
        \begin{align}
          \mathcal{W}(x_1, x_2, n_1, n_2) = \frac{w_1(x_1, n_1, n_2)w_2(x_2, n_1, n_2)}{w(n_1,n_2)}. \label{Eq:JPDAOAltForm}
        \end{align}
    Here $w_1(x_1,n_1,n_2)\geq0$, $w_2(x_2,n_1,n_2)\geq0$, and $w(n_1,n_2)\geq0$ are joint probability distributions for the observables presented by their arguments.  
    In particular, this implies that the conditions
        \begin{align}
            & w(n_1,n_2) = \int_{\mathbb{R}} \D x\, w_i(x,n_1,n_2),\,\,\, i=1,2, \label{Eq:wCondition1} \\
            & \sum_{n_1,n_2=0}^1 w_i(x,n_1,n_2) \delta_{n, n_j} = \mathcal{P}(x,n|\varphi_i, \gamma_j), \label{Eq:wCondition2}
         \end{align}
    are fulfilled and the probabilities $\mathcal{P}(x,n|\varphi,\gamma)$ are indeed marginals of the JPDAO $\mathcal{W}(x_1, x_2, n_1, n_2)$, as stated by Eq. (\ref{Eq:FineLRM}).
        
    Let us specify a particular form of the non-negative functions $w_1(x_1,n_1,n_2)$, $w_2(x_2,n_1,n_2)$, and $w(n_1,n_2)$.
    We chose them such that
         \begin{align}
             & w_1(x,0,0) = \varkappa m(x,\varphi_1) + (1-\varkappa)M(x,\varphi_1), \label{Eq:w1_00} \\
             & w_2(x,0,0) = \varkappa M(x,\varphi_2) + (1-\varkappa)m(x,\varphi_2), \label{Eq:w2_00}
         \end{align}
    where $m(x,\varphi_1)$ and $M(x,\varphi_1)$ are given by Eqs.~(\ref{Eq:mExpression}) and (\ref{Eq:MExpression}), respectively.
    In order to specify the parameter $\varkappa$, we successively substitute Eqs.~(\ref{Eq:w1_00}) and (\ref{Eq:w2_00}) into Eq.~(\ref{Eq:wCondition1}) and equate the results.
    This yields
        \begin{align}
             \varkappa \langle m \rangle_{\varphi_1}&  + (1-\varkappa) \langle M \rangle_{\varphi_1} \nonumber \\
             & = \varkappa \langle M \rangle_{\varphi_2} + (1-\varkappa) \langle m \rangle_{\varphi_2}. \label{:Eq:xiEquation}
         \end{align}
	Solving this equation with respect to $\varkappa$, we obtain the expression
          \begin{align}
              & \varkappa = \left(1 + \frac{\langle M \rangle_{\varphi_2} - \langle m \rangle_{\varphi_1}}{\langle M \rangle_{\varphi_1} - \langle m \rangle_{\varphi_2}}\right)^{-1}, \label{Eq:xiSolution}
          \end{align}
     which specifies the parameter $\varkappa$.			
		
	Fulfillment of Bell inequalities in the form (\ref{Eq:PironioCondition1}) and (\ref{Eq:PironioCondition2}) implies that $\varkappa\in[0,1]$.
	Thus, as it follows from Eqs. (\ref{Eq:w1_00}) and (\ref{Eq:w2_00}), $w_1(x, 0, 0)$ and $w_2(x, 0, 0)$ are convex combinations of non-negative quantities $m(x,\varphi_1)$, $M(x,\varphi_1)$ and $m(x,\varphi_2)$, $M(x,\varphi_2)$, respectively, which guarantees their non-negativity.
	Other values of the probability distributions $w_1(x_1,n_1,n_2)$ and $w_2(x_2,n_1,n_2)$ can be constructed as
		\begin{align}
			w_i(x,1,0) &= \mathcal{P}(x,0|\varphi_i,\gamma_2) - w_i(x,0,0), \label{Eq:w10}\\
			w_i(x,0,1) &= \mathcal{P}(x,0|\varphi_i,\gamma_1) - w_i(x,0,0), \label{Eq:w01}\\
			w_i(x,1,1) &= \mathcal{P}(x|\varphi_i) - \mathcal{P}(x,0|\varphi_i,\gamma_1)\nonumber\\
			&- \mathcal{P}(x,0|\varphi_i,\gamma_2)  + w_i(x,0,0). \label{Eq:w11}
		\end{align}  
	As it is shown in Appendix~\ref{App:ABC}, they are non-negative if $\varkappa\in[0,1]$, i.e., in all cases when inequalities (\ref{Eq:w1_00}) and (\ref{Eq:w2_00}) are satisfied.
	\end{proof}
    
    Nontrivial discretization of the continuous variables $x_i$, corresponding to the special choice of the test function (\ref{Eq:HybridSchemeLambdas}), defines the generalized CHSH inequalities.
    The fact that $x$ belongs or does not belong to the subset $X_i^{(k)}$ can be interpreted as ascribing dichotomic values to the quadrature, see Appendix~\ref{App:CHSH}.
    With our approach, we determine these subsets for each behavior $\mathcal{P}(x,n|\varphi_i, \gamma_j)$ but not an infinite set of tight Bell inequalities, involving continuous variables.


\section{Example: Two-mode squeezed vacuum state}
\label{Sec:TMSVS}

	In this section we consider an application of the developed approach to the two-mode squeezed vacuum state (TMSVS) given by 
	    \begin{align}\label{Eq:TMSVS}
	        |r\rangle=\frac{1}{\cosh{r}}\sum\limits_{n=0}^{\infty}\left(-\tanh{r}\right)^{n}\ket{n}_\mathrm{A}\otimes\ket{n}_\mathrm{B}.
		\end{align}
	Here $r\geq0$ is the squeezing parameter and $\ket{n}_\mathrm{A(B)}$ is a Fock state for the $A(B)$ mode. 
	Experimental procedures for generating this state are widely discussed in literature; see, e.g., Refs.~\cite{Heidmann1987,Dutt2015}.
	
	For our purpose, some remarkable properties of this state are of importance: 
	(i) this is an entangled state; (ii) this state has a positive Wigner function in the form of a Gaussian distribution; (iii) this state is nonclassical since its P function is not positive semidefinite; and (iv) local reduced states for each mode are classical thermal states.
	Two latter properties imply that this state can manifest nonclassical correlations of radiation and no local nonclassicality.
	
	Bell nonlocality for this state without its modification by non-Gaussian operations cannot be verified with balanced homodyne detection alone \cite{nha2004,Garcia2004}.
	Indeed, since its Wigner function and the POVM symbols for $s_\mathrm{A(B)}=0$ [cf. Eq.~(\ref{Eq:POVM_BHD})] are non-negative, a local realistic model is given by phase-space functions, such as in Eq.~(\ref{Eq:PSLRM}).
	However, for $s_\mathrm{A(B)}\geq 0$, nonclassical correlations of radiation with such a measurement layout are still possible.
	Particularly, such measurement outcomes cannot be simulated by statistical mixtures of coherent states. 
	Bell nonlocality for this state can be verified if both modes are analyzed with unbalanced homodyne detection \cite{banaszek98,banaszek1999,banaszek2002}.
	Consequently, nonclassical correlations of radiation in such a configuration exist as well.  	
		
	Here we consider the hybrid measurement scheme considered in Sec.~\ref{Sec:Hybrid}.
	The corresponding behavior in this case is given by
	    \begin{align}\label{Eq:TMSVSBehavior}
	         &\mathcal{P}(x,n|\varphi,\gamma)=n\frac{\exp\left[-\frac{x^2}{\cosh 2r}\right]}{\sqrt{\pi\cosh 2r}}\\\nonumber
	         &+\frac{\exp\left\{-\frac{|\gamma|^2}{\cosh^2 r} -\left[x +\sqrt{2}{\rm Re}(\gamma e^{i\varphi})\tanh r\right]^2\right\}}{(-1)^n\sqrt{\pi}\cosh^2 r}.
	    \end{align}
    First, we will demonstrate that for such a configuration there exist a local realistic model.
    Hence, Bell inequalities (\ref{Eq:BellIneq}) in this case cannot be violated.
    However, inequalities (\ref{Eq:PSBell}) are violated for all appropriate values of the ordering parameters, including the case of $s_\mathrm{A}=1$ and $s_\mathrm{B}=0$.
    Hence, Eq.~(\ref{Eq:PSLRM}) does not have non-negative solutions, and the partial solution (\ref{Eq:DevDep}) is not non-negative.
    
    \subsection{Local realistic model}   

	In this section we prove that a non-negative \mbox{JPDAO} can always be constructed for the behavior~(\ref{Eq:TMSVSBehavior}).
    It can be shown analytically that inequalities (\ref{Eq:PironioCondition1}) and (\ref{Eq:PironioCondition2}) are fulfilled for any measurement settings $\gamma_1$, $\gamma_2$ and $\varphi_1$, $\varphi_2$ if the squeezing parameter $r\gtrsim 1.66$.
    In order to demonstrate this, we rewrite Eq.~(\ref{Eq:mExpression}) as
          \begin{align}\label{Eq:m}
              m(x,\varphi) &= \mathcal{P}(x|\varphi) \max \{ \mathcal{P}(0|\gamma_1; x;\varphi)  \nonumber \\
                & + \mathcal{P}(0|\gamma_2; x;\varphi) - 1, 0 \},
          \end{align}
     where we have used Bayes' formula $\mathcal{P}(x,0|\varphi,\gamma) = \mathcal{P}(x|\varphi)\mathcal{P}(0|\gamma;x;\varphi)$.
     The conditional probability $\mathcal{P}(0|\gamma;x;\varphi)$ in the case of the TMSVS reads as
          \begin{align}
              & \mathcal{P}(0|\gamma; x;\varphi) = \frac{\sqrt{\cosh 2r}}{\cosh^2 r} \nonumber \\
              & \times\exp \left\{ -\frac{\cosh 2r}{\cosh^2 r}\left[\Re \left(\gamma e^{-i\varphi}\right) + \frac{x}{\sqrt{2}}\tanh 2r\right]^2\right\} \nonumber \\
              & \times \exp \left\{ -\frac{\left[\Im \left(\gamma e^{-i\varphi}\right)\right]^2}{\cosh^2 r} \right\}.
           \end{align}
     This probability cannot exceed its maximum  $\sqrt{\cosh{2r}}/\cosh^2 r$.
     As it follows from Eq. (\ref{Eq:m}), one gets $m(x,\varphi) = 0$ for any $\varphi$, $x$ and for the values of $r$ satisfying the condition 
           \begin{align}\label{Eq:CondMax}
               \sqrt{\cosh{2r}}/\cosh^2 r \leq 1/2.
           \end{align}
     This implies that Eqs. (\ref{Eq:PironioCondition1}) and (\ref{Eq:PironioCondition2}) take the form
           \begin{align}
               0 \leq \langle M_i \rangle,
           \end{align}
     where $i = 1, 2$.
     This inequality is always fulfilled due to the construction  of $\langle M_i \rangle$. 
     Therefore, for $r$ satisfying the condition (\ref{Eq:CondMax}), i.e., for $r\geq\arcosh\sqrt{2\sqrt{3}+4}\approx 1.66$, a non-negative JPDAO always exists. 
        
     In order to confirm the local character of the behavior (\ref{Eq:TMSVSBehavior}) for other values of $r$, we use the technique of numerical optimization.
     The corresponding optimization problem can be formulated via rewriting Eqs. (\ref{Eq:PironioCondition1}) and (\ref{Eq:PironioCondition2}) as
          \begin{align}
              \label{Eq:OptimizationInequality}F(\varphi_1,\varphi_2, \gamma_1, \gamma_2;r) = \langle m_1 \rangle - \langle M_2 \rangle \leq 0.
          \end{align}
     If the maximum value of $F(\varphi_1,\varphi_2, \gamma_1, \gamma_2;r)$ is positive, the inequality (\ref{Eq:OptimizationInequality}) is violated for the corresponding values of $\varphi_1$, $\varphi_2$, $\gamma_1$, $\gamma_2$, and $r$, and, consequently, the behavior is nonlocal.
     Similarly, if the maximum value of $F(\varphi_1,\varphi_2, \gamma_1, \gamma_2;r)$ is non-negative, the behavior does not exhibit nonlocality in the aforementioned scenario.
     Applying the simplicial homology global optimization method \cite{SHGO}, we obtain that the maximum value of $F(\varphi_1,\varphi_2, \gamma_1, \gamma_2;r)$ is exactly zero.
     This implies that the considered hybrid measurement scheme with the TMSVS can be treated within a local realistic model.
     In particular, this means that Bell inequalities (\ref{Eq:BellIneq}) in the given scenario cannot be violated.
        
     The JPDAO (\ref{Eq:JPDAOAltForm}) can also be presented in the form of Eq.~(\ref{Eq:GenSolution}), where $\mathcal{W}_\mathrm{h}$ and  $\mathcal{W}_\mathrm{p}$ are given by Eqs.~(\ref{Eq:HomogeneousSolution}) and (\ref{Eq:DevDep}), respectively.
     In this case the functions $C(x_1,x_2)$ and $C_{ij}(x_1,x_2)$ are reduced to the form
        \begin{align}
            & C(x_1,x_2) = \mathcal{H}(x_1,x_2,1,1), \\
            & C_{00}(x_1,x_2) = \mathcal{H}(x_1,x_2,0,0) - C(x_1,x_2), \\
            & C_{01}(x_1,x_2) = \mathcal{H}(x_1,x_2,0,1) + C(x_1,x_2), \\
            & C_{10}(x_1,x_2) = \mathcal{H}(x_1,x_2,1,0) + C(x_1,x_2). 
        \end{align}
     Here the function 
        \begin{align}
            & \mathcal{H}(x_1,x_2,n_1,n_2) \nonumber \\
            & = \mathcal{W}(x_1,x_2,n_1,n_2) - \mathcal{W}_{\mathrm{p}}(x_1,x_2,n_1,n_2),
        \end{align}
    is obtained as a difference of the function (\ref{Eq:JPDAOAltForm}) and the particular solution (\ref{Eq:DevDep}).

\subsection{Nonclassical correlations of radiation}

    Despite that the hybrid measurement scheme discussed in Sec.~\ref{Sec:Hybrid} cannot be used for testing nonlocal properties of the TMSVS, it is still capable for testing nonclassical correlations of radiation.
    This implies that no non-negative solutions $P(\alpha_{\mathrm{A}},\alpha_{\mathrm{B}};s_{\mathrm{A}},s_{\mathrm{B}})$ of Eq. (\ref{Eq:PSLRM}) exist for all appropriate values of $s_{\mathrm{A}}=1$ and $s_{\mathrm{B}}\in[0,1]$.
    In this section we will prove this fact via demonstrating violations of Bell-like inequality (\ref{Eq:PSBell}) for the scenario of the TMSVS analyzed with the hybrid measurement scheme.
    
    Let us choose the test function $\lambda$ in inequality (\ref{Eq:PSBell}) as
    	\begin{align}\label{Eq:PSLambda}
    	    \lambda(x,n,\varphi,\gamma) = \delta(x-x_0) \delta_{\varphi, \varphi_0}\left( \delta_{n,0} \chi(\gamma;\alpha_0) - \frac{D}{2}\right),
    	\end{align}
    where $\delta(x)$ and $\delta_{n,k}$ are the Dirac delta function and the Kronecker delta, respectively.
    In particular, such a choice means that we can consider only a single local-oscillator phase at Bob's side.
    The coefficient $D$ is given by
    	\begin{align}
    		D = \sup_{\alpha} \sum_{j=1}^2 \chi(\gamma_j;\alpha_0) \Pi(0|\gamma_j;\alpha).\label{Eq:ConstantC}
    	\end{align}
    The argument $\gamma$ of the function $\chi(\gamma;\alpha_0)$ takes only two real values, $\gamma\in\{\gamma_1,\gamma_2\}$.
    The corresponding values of the function are chosen as
        \begin{align}
            & \chi(\gamma_1;\alpha_0) = -(\alpha_0 - \gamma_2) e^{-|\alpha_0 - \gamma_2|^2}, \\
            & \chi(\gamma_2;\alpha_0) = (\alpha_0 - \gamma_1) e^{-|\alpha_0 - \gamma_1|^2}. 
        \end{align}
    The real constants $x_0$ and $\alpha_0$ are adjusted to optimize violations of inequality (\ref{Eq:PSBell}), and $\varphi_0$ is an arbitrary phase.
    
    The left-hand side of inequality (\ref{Eq:PSBell}) with the chosen function (\ref{Eq:PSLambda}) is reduced to the form
        \begin{align}\label{Eq:LHSPSBell}
			 &\sum_{i,j=1}^2 E(\lambda|\varphi_i,\gamma_j) \\
			 &= \sum_{j=1}^2 \chi(\gamma_j;\alpha_0) \mathcal{P}(x_0,0|\varphi_0,\gamma_j) - D \mathcal{P}(x_0|\varphi_0).  \nonumber
		\end{align}    
    Here $\mathcal{P}(x_0,0|\varphi_0,\gamma_j)$ is the particular value of the behavior and $\mathcal{P}(x_0|\varphi_0)$ is the marginal distribution for the quadrature given by Eq.~(\ref{Eq:DistrX}).
    It can be directly checked that the right-hand side of inequality (\ref{Eq:PSBell}) with the function (\ref{Eq:PSLambda}) is zero:
          \begin{align}
			  \label{Eq:PSBellRHS}& \sup_{\boldsymbol{\alpha}} \sum_{i,j=1}^2 E(\lambda|\varphi_i,\gamma_j;\boldsymbol{\alpha};0,1) =0.
		  \end{align}
  	Therefore, Bell-like inequality (\ref{Eq:PSBell}) is reduced to the form
  			\begin{align}\label{Eq:PSBellSpecial}
  				\sum_{j=1}^2 \chi(\gamma_j;\alpha_0) \mathcal{P}(x_0,0|\varphi_0,\gamma_j) \leq D \mathcal{P}(x_0|\varphi_0).
  			\end{align}
  	Both sides of this inequality can be directly sampled from the measurement data.
  	It is also important that this inequality does not depend on the ordering parameter $s_\mathrm{A}$.
  	Hence, its violation implies both facts: the impossibility of simulation of the measurement outcomes with statistical mixtures of coherent states and the impossibility of more general simulations with any $s$-parameterized phase-space function. 
  	
\begin{figure}[ht!]
                \includegraphics[width=1\linewidth]{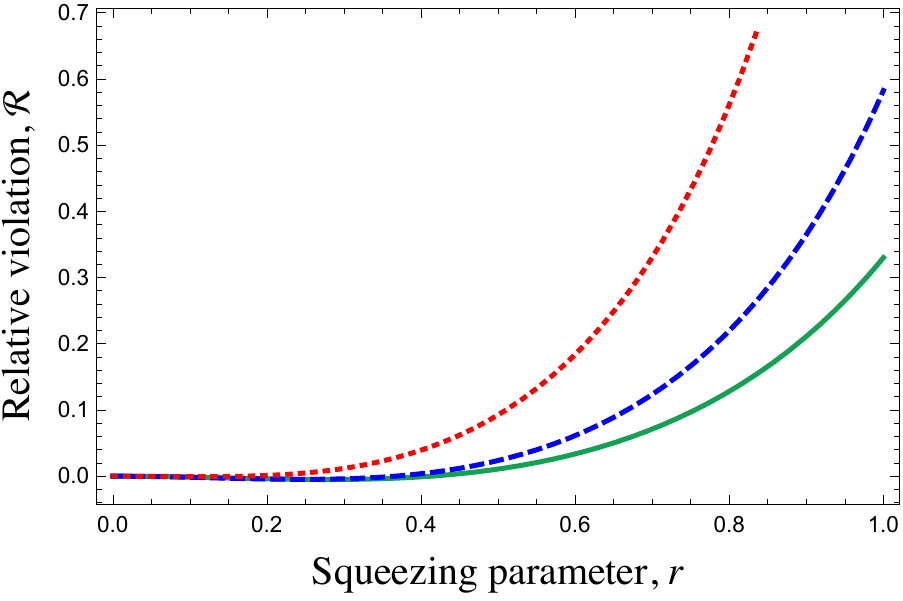}
                \caption{\label{Fig:rPlot} The relative violation $\mathcal{R}$, cf. Eq.~(\ref{Eq:R}), of inequality~(\ref{Eq:PSBellSpecial}) is shown as a function of the squeezing parameter $r$ for the TMSVS (\ref{Eq:TMSVS}).
                The measurement-device settings are chosen as $\varphi_0=0$, $\gamma_1 = 0$, $\gamma_2 = 1$.
                Solid, dashed, and dotted lines correspond to the detection efficiencies ($\eta_A = 0.7$, $\eta_B = 0.6$), ($\eta_A = 0.7$, $\eta_B = 0.7$), and ($\eta_A = 0.8$, $\eta_B = 0.7$), respectively.}
        \end{figure}

  	We consider the relative violation of inequality (\ref{Eq:PSBellSpecial}) as the ratio of difference of its left- and right-hand sides to the latter:
	  		\begin{align}\label{Eq:R}
	  		 	\mathcal{R}=\frac{\sum\limits_{j=1}^2 \chi(\gamma_j;\alpha_0) \mathcal{P}(x_0,0|\varphi_0,\gamma_j)-D \mathcal{P}(x_0|\varphi_0)}{D \mathcal{P}(x_0|\varphi_0)}.	
	  		\end{align}
  	It is shown in Fig.~\ref{Fig:rPlot} as a function of the squeezing parameter $r$ for different values of the detection efficiencies.
  	The parameters $x_0$ and $\alpha_0$ are chosen to maximize the relative violation and fulfill  the condition $\mathcal{P}(x|\varphi) > 0.1/\sqrt{\pi \cosh{2r}}$.
  	Here we consider different detection efficiencies.
  	These efficiencies are included in the behavior (\ref{Eq:TMSVSBehavior}) as it is discussed in Appendix~\ref{App:BehEta}.
  	Violations can be clearly verified for accessible conditions of experiments.
  	This means that despite the existence of a local realistic model, the obtained correlations cannot be simulated with statistical mixture of coherent states and, more generally, with phase-space functions.
  	
  	Testing nonclassical correlations of radiation with Bell-like inequalities (\ref{Eq:PSBell}) is based on a finite number of measurement-device settings.
  	Hence, with our method we check nonclassical correlations for behaviors. 
  	Alternative approaches \cite{Agudelo2013,Koehnke2021} deal with tomographic reconstruction of the regularized P function \cite{Kiesel11a} and test nonclassical correlations for quantum states.

\section{Example: Schr\"odinger-Cat state}
\label{Sec:CatState}

    In this section we introduce an example, leading to violations of local realism for the hybrid measurement scheme.
    For this purpose, we consider a hybrid entangled state \cite{jeong14} also known as a Schr\"odinger-cat state \cite{Agudelo2017} given by 
        \begin{align}
            \label{Eq:CatState}
            \ket{\Psi} = \frac{1}{\sqrt{2}} \Big( \ket{\alpha_0}_\mathrm{A} \otimes \ket{0}_\mathrm{B} + \ket{-\alpha_0}_\mathrm{A} \otimes \ket{1}_\mathrm{B} \Big),
        \end{align}
    where $\ket{\alpha_0}$, $\ket{0}$, and $\ket{1}$ indicate coherent, vacuum, and single photon Fock states, respectively.
    The generation scheme of this state has been discussed in Refs.~\cite{jeong14,Agudelo2017}.
    
    Let Alice and Bob perform balanced and unbalanced homodyning with detection efficiencies $\eta_\mathrm{A}$ and $\eta_\mathrm{B}$, respectively.
    The corresponding behavior reads
        \begin{align}
            \label{Eq:CatStateBehavior}
            & \mathcal{P}(x, 0|\varphi, \gamma) = \frac{e^{-|\gamma|^2}}{2}\left\{ \vphantom{\left[ \gamma \exp\left(- (x - i \sqrt{2 \eta_A} \Im \alpha_0 e^{-i\varphi})^2\right) \right]}
            \Pi(x| \varphi; \sqrt{\eta_\mathrm{A}}\alpha_0; -1) \right. \nonumber\\
            & \left. + \Pi(x| \varphi; -\sqrt{\eta_\mathrm{A}}\alpha_0; -1)(1 - \eta_\mathrm{B} + \eta_\mathrm{B} |\gamma|^2) \right. \nonumber\\
            & \left. + \frac{2}{\sqrt{\pi}}\sqrt{\eta_B} \exp\left(-2|\alpha_0|^2\right) \right. \nonumber \\
            & \left. \times \Re \left[ \gamma \exp\left(- (x - i \sqrt{2 \eta_A} \Im \alpha_0 e^{-i\varphi})^2\right) \right] \right\}.
        \end{align} 
    Here
        \begin{align}\label{Eq:CatStateMarginalBehavior}    
            \mathcal{P}(x|\varphi) = & \frac{1}{2}\left(\Pi(x| \varphi, \sqrt{\eta_A}\alpha_0; -1) \right. \nonumber \\
            & \left. + \Pi(x| \varphi; -\sqrt{\eta_A}\alpha_0; -1) \right),
        \end{align}
    is the marginal behavior at the Alice side, and $\Pi(x|\varphi;\alpha;-1)$ is given by Eq.~(\ref{Eq:POVM_BHD}) for $(-s)=-1$.
    
\begin{figure}[ht!]
            \includegraphics[width=1\linewidth]{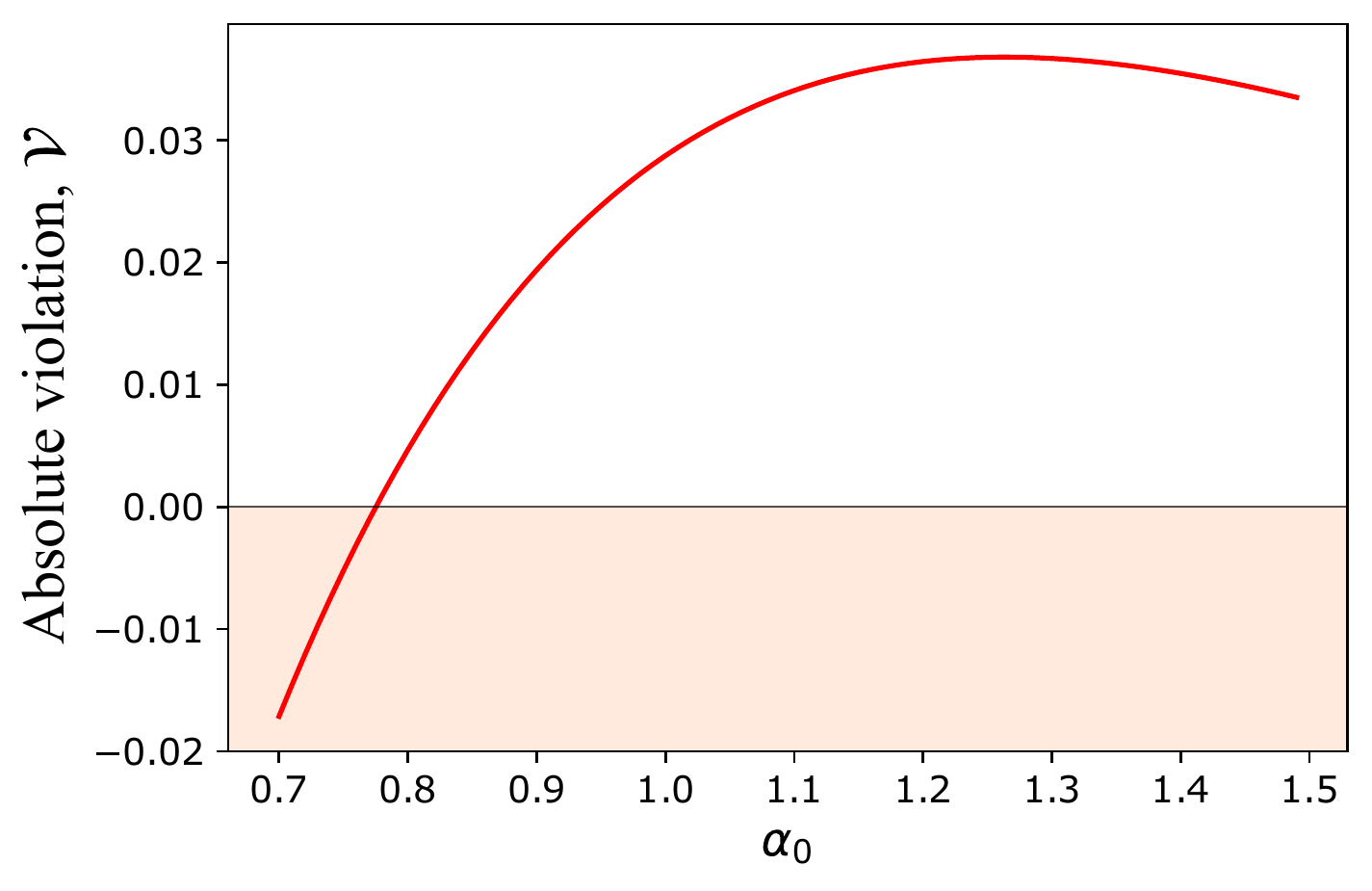}
            \caption{\label{Fig:CatStatePlot} The absolute violation $\mathcal{V}$, cf. Eq.~(\ref{Eq:AbsoluteViolation}), of inequalities (\ref{Eq:PironioCondition1}) and (\ref{Eq:PironioCondition2}) are shown as a function of $\alpha_0$ for $\Im\alpha=0$.
            The phases of local oscillator at Alice's side are chosen as $\varphi_1 = 0$, $\varphi_2 = \frac{\pi}{2}$.
            The values of phase-space displacement at Bob's side are $\gamma_1 = \frac{i}{4}$, $\gamma_2 = -\frac{i}{4}$.
            The detection efficiencies have the values $\eta_\mathrm{A} = 0.95$, $\eta_\mathrm{B} = 0.95$.
            The shaded area indicates the negative values of $\mathcal{V}$, for which the corresponding behavior is local realistic.}
        \end{figure}

    We consider the absolute violation of inequalities (\ref{Eq:PironioCondition1}) and (\ref{Eq:PironioCondition2}),
        \begin{align}
            \label{Eq:AbsoluteViolation}
            \mathcal{V} = \max\left\{ \langle m \rangle_{\varphi_1} - \langle M \rangle_{\varphi_2}, \langle m \rangle_{\varphi_2} - \langle M \rangle_{\varphi_1} \right\},
        \end{align}
    as a function of $\alpha_0$ for $\Im \alpha_0=0$.
    The corresponding dependence is shown in Fig.~\ref{Fig:CatStatePlot} for particular settings at both sides.
    One can observe, that nonlocality can be indicated for $\alpha_0\gtrsim 0.8$.

\section{Conclusions}
\label{Sec:Conclusions}

	To conclude, we have studied a fundamental link between Bell nonlocality and nonclassical correlations of radiation.
	Both phenomena are described within the framework of one mathematical formulation.
	Indeed, the existence of a local realistic model implies that a system of linear inhomogeneous equations for the JPDAO has at least one non-negative solution.
	The right-hand side of this system is given by the behavior---the probability distribution for pairs of measurement outcomes directly sampled from the experimental data.
	A particular solution to this system is naturally expressed via phase-space quasiprobability distributions and phase-space symbols of the POVMs.
	The impossibility to classically simulate the measurement outcomes with statistical mixtures of coherent states or, more generally, with phase-space functions implies that this solution has negative values.
	Negativity of the considered particular solution does not mean the nonexistence of another solution, which can be non-negative.
	For this reason, nonclassical correlations of radiation can be feasible even in the cases when measurement outcomes are explained within a local realistic model.   

	We have demonstrated that Bell nonlocality and nonclassical correlations of radiation can also be checked with a similar mathematical tools.
	The tests of these phenomena are expressed in terms of inequalities.
	In the case of testing local realism, these are Bell inequalities, which can be violated if the behavior is nonlocal.
	In the case of testing nonclassical correlations of radiation, the right-hand side of Bell inequalities should be  modified.   
	Our inequalities are naturally formulated for any type of measurement outcomes, including the case of continuous variables and hybrid measurement schemes involving both discrete and continuous outcomes.  

	Our results are demonstrated within an example of the TMSVS.
	These states demonstrate nonclassical correlations of radiation and no local nonclassicality.
	It is known that Bell nonlocality for these states can be verified with unbalanced homodyne detection and cannot be verified with balanced homodyne detection.
	We have analyzed a hybrid scheme involving both types of the measurements.
	It has been shown that the corresponding measurement data can also be explained with a local realistic model.
	However, they cannot be simulated by a statistical mixture of coherent states or by phase-space functions. 
	
	\begin{acknowledgments}
		S.R. and A.A.S. thank K. Pregracke for fruitful discussions. W.V. and A.A.S. would also like to thank J.~Sperling for his valuable comments on the manuscript.
		V.S.K., I.S.Y., and A.A.S. acknowledge support from the National Research Foundation of Ukraine through the Project No. 2020.02/0111, ``Nonclassical and hybrid correlations of quantum systems under realistic conditions.''
	\end{acknowledgments}


\appendix

\section{Phase-space functions}
\label{App:PSF}
	In this Appendix we briefly refer to relations between operators and phase-space functions \cite{cahill69,cahill69a}.
	For two optical modes, the Cahill-Glauber quasiprobability distribution parametrized by two ordering parameters $s_\mathrm{A}$ and $s_\mathrm{B}$ is expressed via the density operator $\hat{\rho}$ as  
		\begin{align}
			P(\alpha_\mathrm{A},\alpha_\mathrm{B};s_\mathrm{A},s_\mathrm{B})=\Tr\left[\hat{\rho}\hat{P}(\alpha_\mathrm{A};s_\mathrm{A})\otimes\hat{P}(\alpha_\mathrm{B};s_\mathrm{B})\right].
		\end{align}
	Similarly, the phase-space symbols of the POVMs are related to the POVMs as
		\begin{align}
			&\Pi_{\mathrm{A}}(A|a;\alpha_\mathrm{A};s_\mathrm{A})=\frac{1}{\pi}\Tr\left[\hat{\Pi}_\mathrm{A}(A|a)\hat{P}(\alpha_\mathrm{A};s_\mathrm{A})\right],\\
			&\Pi_{\mathrm{B}}(B|b;\alpha_\mathrm{B};s_\mathrm{B})=\frac{1}{\pi}\Tr\left[\hat{\Pi}_\mathrm{B}(B|b)\hat{P}(\alpha_\mathrm{B};s_\mathrm{B})\right].
		\end{align} 
	In these relations, the operator kernel reads
		\begin{align}
			\hat{P}(\alpha;s)=
			\frac{2}{\pi(1-s)}:\exp\left[-\frac{2}{1-s}(\hat{a}^\dag-\alpha^\ast)(\hat{a}-\alpha)\right]:,
		\end{align}
	where $\hat{a}$ and $\hat{a}^\dag$ are field annihilation and creation operators, respectively.

\section{Standard and generalized CHSH inequalities}
\label{App:CHSH}    
    In this Appendix we analyze two issues related to CHSH inequalities \cite{CHSH}.
    First, we refer to relations between the general form (\ref{Eq:BellIneq}) of Bell inequalities and their particular case---CHSH inequalities.
    Second, we analyze the test function (\ref{Eq:HybridSchemeLambdas}) in order to demonstrate its connection with a generalization of CHSH inequalities.
    
    The standard Bell-CHSH scenario is a particular case of the scheme depicted in Fig.~\ref{Fig:Bell} with $A_i$, $B_j$ taking two discrete values, say $0$ and $1$.
    We consider four CHSH inequalities indexed by numbers $m=a_1,a_2$ and $n=b_1,b_2$.
    The corresponding four test functions are given by
        \begin{align}
            \lambda_{mn}(A,B,a,b)=(2\delta_{A,B}-1)(1-2\delta_{a,m}\delta_{b,n}).
        \end{align}
    In this case, the expected value (\ref{Eq:ExpectedLambda}) is given by
        \begin{align}
            E(\lambda_{mn}&|a_i,b_j)=\left[1-2\delta_{a_i,m}\delta_{b_j,n}\right]\nonumber\\
            &\times\left[P(\textrm{same}|a_i,b_j)-P(\textrm{different}|a_i,b_j)\right],
        \end{align}
    where
        \begin{align}
            P(\textrm{same}&|a_i,b_j)\nonumber\\
            &=P(0,0|a_i,b_j)+P(1,1|a_i,b_j),
        \end{align}
    and    
        \begin{align}
            P(\textrm{different}&|a_i,b_j)\nonumber\\
            &=P(0,1|a_i,b_j)+P(1,0|a_i,b_j)
        \end{align}
    are probabilities to get the same and different values of $A$ and $B$, respectively.
    The right-hand side of Eq.~(\ref{Eq:BellIneq}) for this test function,
        \begin{align}
            \sup_{\mathbf{A}, \mathbf{B}} \sum_{i,j} \lambda_{mn}(A_i,B_j,a_i,b_j)=2.
        \end{align}
     This yields the set of four CHSH inequalities,
        \begin{align}
        	\label{Eq:CHSH}
            \sum\limits_{i,j=1}^2E(\lambda_{mn}|a_i,b_j)\leq 2.
        \end{align}
      The other four inequalities can be obtained by changing the sign of the chosen test function $\lambda_{mn}(A,B,a,b)$.
      
	In order to show the relation between standard and generalized CHSH inequalities, we rewrite the latter ones in terms of dichotomized quadratures.
	Let us start with the test function $\lambda^{(1)}(x,n|\varphi,\gamma)$ from Eq. (\ref{Eq:HybridSchemeLambdas}).
	The corresponding Bell inequality can be expressed as
		\begin{align}
			\label{Eq:DichotomizedHybridInequality}
			& \tilde{\mathcal{P}}(0,0|\varphi_1, \gamma_1) - \tilde{\mathcal{P}}(0,1|\varphi_1, \gamma_2) \nonumber \\
			& -\tilde{\mathcal{P}}(0,0|\varphi_2,\gamma_1) - \tilde{\mathcal{P}}(1,0|\varphi_2,\gamma_2) \leq 0,
		\end{align}
	where 
		\begin{align}
			\tilde{\mathcal{P}}(\tilde{x},n|\varphi_i, \gamma_j) = & (1 - \tilde{x}) \int_{X^{(1)}_i} \D x\, \mathcal{P}(x,n|\varphi_i,\gamma_j) \nonumber \\ 
			& + \tilde{x} \int_{\mathbb{R} \setminus X^{(1)}_i} \D x\, \mathcal{P}(x,n|\varphi_i,\gamma_j)
		\end{align}
	is the joint probability distribution of a dichotomized quadrature $\tilde{x}(\varphi_i) \in \{0, 1\}$, $i = 1, 2$, given by
		\begin{align}
			\label{Eq:DichotomizedQuadrature}
			\tilde{x}(\varphi_i) = 1 - \mathbf{I}\left(x;X^{(1)}_i\right)
		\end{align}
	and $n(\gamma)$.
	In fact, Eq. (\ref{Eq:DichotomizedHybridInequality}) is the CHSH inequality associated with $\lambda_{a_2 b_1}(A,B,a,b)$.
	One can verify this by expressing the 16 probabilities $\mathcal{P}(A,B|a,b)$ present in Eq.~(\ref{Eq:CHSH}) via eight linearly independent ones, $\mathcal{P}(0,0|a_1,b_1)$, $\mathcal{P}(0,0|a_1,b_2)$, $\mathcal{P}(0,1|a_1,b_2)$, $\mathcal{P}(0,0|a_2,b_1)$, $\mathcal{P}(1,0|a_2,b_1)$, $\mathcal{P}(0,0|a_2,b_2)$, $\mathcal{P}(1,0|a_2,b_2)$, $\mathcal{P}(0,1|a_2,b_2)$:
		\begin{align}
			\mathcal{P}(0,1|a_1,b_1) = & \mathcal{P}(0,0|a_1,b_2) + \mathcal{P}(0,1|a_1,b_2) \nonumber \\ 
			& - \mathcal{P}(0,0|a_1,b_1), \\
			\mathcal{P}(1,B|a_1,b_j) = & \mathcal{P}(0,B|a_2,b_j) + \mathcal{P}(1,B|a_2,b_j) \nonumber \\
			& - \mathcal{P}(0,B|a_1,b_j), \\
			\mathcal{P}(0,1|a_2,b_1) = & \mathcal{P}(0,0|a_2,b_2) + \mathcal{P}(0,1|a_2,b_2) \nonumber \\
			& - \mathcal{P}(0,0|a_2,b_1), \\
			\mathcal{P}(1,1|a_2,b_j) = & 1 - \mathcal{P}(0,0|a_2,b_2) - \mathcal{P}(0,1|a_2,b_2) \nonumber \\
			& - \mathcal{P}(1,0|a_2,b_j),
		\end{align}	
	where $B = 0, 1$ and $j = 1, 2$.
    For the remaining test function $\lambda^{(2)}(A,B,a,b)$,
	one can similarly introduce dichotomized quadrature by replacing $X^{(1)}_i$ with $X^{(2)}_i$ in Eq. (\ref{Eq:DichotomizedQuadrature}).

\section{Asymmetric Bell correlations}
\label{App:ABC}
    In this Appendix we show that particular values of the probability distributions $w_1(x_1, n_1, n_2)$ and $w_2(x_2, n_1, n_2)$, cf. Eqs. (\ref{Eq:w10}), (\ref{Eq:w01}), and (\ref{Eq:w11}), are non-negative if the parameter $\varkappa$, defined by Eq. (\ref{Eq:xiSolution})  belongs to the interval $[0,1]$.
    Without loss of generality, let us consider $w_1(x,1,0)$.
    The restriction $w_1(x,1,0) \geq 0$ yields
        \begin{align}
            \mathcal{P}(x,0|\varphi_1, \gamma_2) \geq w_1(x,0,0). \label{Eq:w1_00Condition}
        \end{align}
    Note that if $\varkappa \in [0,1]$, then $w_1(x,0,0) \in [m(x,\varphi_1),M(x,\varphi_1)]$, since this function is a convex combination of the bounds of this interval.
    Equation (\ref{Eq:MExpression}) implies that $\mathcal{P}(x,0|\varphi_1, \gamma_2) \geq m(x,\varphi_1)$.
    Therefore $\mathcal{P}(x,0|\varphi_1, \gamma_2)$ is greater than the upper bound of the interval to which $w_1(x,0,0)$ belongs.
    Thus the condition from Eq. (\ref{Eq:w1_00Condition}) is satisfied.
    Similarly, one can show that $w_1(x,0,1)$, $w_2(x,1,0)$, and $w_2(x,0,1)$ are non-negative as well.
    
    For $w_1(x,1,1)$ and $w_2(x,1,1)$, defined by Eq. (\ref{Eq:w11}), to be non-negative, the inequality
        \begin{align}
            w_i(x,0,0) \geq \mathcal{P}(x,0|\varphi_i,\gamma_1) + 
            \mathcal{P}(x,0|\varphi_i,\gamma_2) - \mathcal{P}(x|\varphi_i) \label{Eq:wi00Condition}
        \end{align}
    must hold, where $i=1,2$.
    Considering that $w_i(x,0,0) \in [m(x,\varphi_i),M(x,\varphi_i)]$ if $\varkappa \in [0,1]$, the right-hand side of Eq. (\ref{Eq:wi00Condition}) is always greater than or equal to $m(x,\varphi_i)$, according to Eq. (\ref{Eq:mExpression}).

\section{Behavior for the hybrid measurement scheme with detector losses}
\label{App:BehEta}    
	
	In this Appendix we introduce the explicit form for a particular value of the behavior (\ref{Eq:TMSVSBehavior}) in the presence of detector losses.
	It is given by
	\begin{widetext}
		\begin{align}
			\mathcal{P}(x,0|\varphi,\gamma)= \frac{1}{\sqrt{\pi \sigma_1 \sigma_3}}\exp{\left[
			-\frac{\sigma_2}{\sigma_1}\left( \Re \gamma e^{-i\varphi} + x\sqrt{\frac{\eta_\mathrm{A} \eta_\mathrm{B}}{2}} \frac{\sinh{2r}}{\sigma_2}  \right)^2\right]}
			\exp{\left[-
			    \frac{\left(\Im \gamma e^{-i\varphi}\right)^2}{\sigma_3}
			\right]}\exp{\left[-\frac{x^2}{\sigma_2}\right]},
		\end{align}
	\end{widetext}
	where
	    \begin{align}
	        & \sigma_1 = \eta_\mathrm{B} \cosh^2 r + (1 - \eta_\mathrm{B})(1 + 2\eta_\mathrm{A} \sinh^2 r), \\
	        & \sigma_2 = 1 + 2\eta_\mathrm{A}\sinh^2 r, \\
	        & \sigma_3 = 1 + \eta_\mathrm{B}\sinh^2 r,
	    \end{align}
	and $\eta_\mathrm{A}$ and $\eta_\mathrm{B}$ are the detection efficiencies for balanced and unbalanced homodyne detection, respectively.
    The marginal distribution (\ref{Eq:DistrX}) in this case reads
    	\begin{align}
    	    \mathcal{P}(x|\varphi) = \frac{1}{\sqrt{\pi \sigma_2}} \exp \left[
    	    -\frac{x^2}{\sigma_2}
    	    \right].
    	\end{align}
    Other values of the behavior do not appear in inequality (\ref{Eq:PSBellSpecial}).	
	
\bibliography{biblio}

\end{document}